\documentclass[12pt]{article}

\usepackage{amsmath,amssymb,amsfonts,amsthm,hyperref,bbm,latexsym,epsfig}
\usepackage{graphicx,multirow}
\usepackage{caption}

\newcommand{\bs}{\begin{subequations}}
\newcommand{\es}{\end{subequations}}
\newcommand{\be}{\begin{equation}}
\newcommand{\ee}{\end{equation}}
\newcommand{\ba}{\begin{eqnarray}}
\newcommand{\ea}{\end{eqnarray}}
\newcommand{\no}{\nonumber \\}

\newcommand{\ab}{\bar{a}}
\newcommand{\bb}{\bar{b}}
\newcommand{\cb}{\bar{c}}
\newcommand{\db}{\bar{d}}
\newcommand{\eb}{\bar{e}}
\newcommand{\fb}{\bar{f}}

\allowdisplaybreaks

\begin{document}

\title{
\normalsize \hfill CFTP/17-009
\\[6mm]
\LARGE Lepton mixing and the charged-lepton \\ mass ratios}

\author{
\addtocounter{footnote}{2}
Darius~Jur\v{c}iukonis$^{(1)}$\thanks{E-mail: \tt darius.jurciukonis@tfai.vu.lt}
\ {\normalsize and}
Lu\'\i s~Lavoura$^{(2)}$\thanks{E-mail: \tt balio@cftp.tecnico.ulisboa.pt}
\\*[3mm]
$^{(1)} \! $
\small University of Vilnius,
\small Institute of Theoretical Physics and Astronomy, \\
\small Saul\.{e}tekio ave.~3, LT-10222 Vilnius, Lithuania 
\\[2mm]
$^{(2)} \! $
\small Universidade de Lisboa, Instituto Superior T\'ecnico, CFTP, \\
\small 1049-001 Lisboa, Portugal
\\*[2mm]
}

\date{\today}

\maketitle

\begin{abstract}
  We construct a class of renormalizable models for lepton mixing
  that generate predictions given in terms of the charged-lepton mass ratios.
  We show that one of those models leads,
  when one takes into account the known experimental values,
  to almost maximal $CP$-breaking phases
  and to almost maximal neutrinoless double-beta decay.
  We study in detail the scalar potential of the models,
  especially the bounds imposed by unitarity
  on the values of the quartic couplings.
\end{abstract}

\newpage

\section{Introduction and notation}

In this paper we present a class of renormalizable models
that are extensions of the Standard Model (SM).
The models have gauge group $SU(2) \times U(1)$ just as the SM.
They feature an extended scalar sector,
with three $SU(2)$ doublets $\phi_k = \left( \phi_k^+,\ \phi_k^0 \right)^T$
($k = 1, 2, 3$)
instead of one;
we let $\tilde \phi_k = \left( {\phi_k^0}^\ast,\, - \phi_k^- \right)^T$
denote the conjugate doublets.
The leptonic sector is also extended,
with the addition to the SM of three right-handed
(\textit{i.e.}~$SU(2)$-singlet)
neutrinos $\nu_{R\psi}$,\footnote{In this paper the Greek letters $\psi$,
  $\alpha$,
  $\beta$,
  and $\gamma$ in general run over the lepton flavours $e$,
  $\mu$,
  and $\tau$.
  Whenever we use $\left( \alpha, \beta, \gamma \right)$
  we mean a permutation of $\left( e, \mu, \tau \right)$,
  \textit{i.e.}~$\alpha \neq \beta \neq \gamma \neq \alpha$.}
which enable a type-I seesaw mechanism~\cite{seesaw}
to suppress the standard-neutrino masses.
Our models have family-lepton-number symmetries
\be
\label{leptonnumbers}
D_{L\psi} \to e^{i \xi_\psi} D_{L\psi}, \quad
\psi_R \to e^{i \xi_\psi} \psi_R, \quad
\nu_{R\psi} \to e^{i \xi_\psi} \nu_{R\psi},
\ee
where the phases $\xi_e$,
$\xi_\mu$,
and $\xi_\tau$ are arbitrary and uncorrelated.
In transformation~\eqref{leptonnumbers},
$\psi_R$ denotes the right-handed charged leptons
and $D_{L\psi} = \left( \nu_{L\psi},\, \psi_L \right)^T$
denotes the $SU(2)$ doublets
of left-handed neutrinos $\nu_{L\psi}$ and charged leptons $\psi_L$.
In our models both the charged-lepton mass matrix $M_\ell$
and the neutrino Dirac mass matrix $M_D$ are diagonal,
because they originate in Yukawa couplings that respect the
family-lepton-number symmetries~\eqref{leptonnumbers}.
Thus,
\be
M_\ell = \mathrm{diag} \left( \ell_e,\, \ell_\mu,\, \ell_\tau \right),
\quad
M_D = \mathrm{diag} \left( D_e,\, D_\mu,\, D_\tau \right),
\ee
where
\be
\label{malpha}
\left| \ell_\psi \right| = m_\psi
\ee
are the charged-lepton masses.
The symmetries~\eqref{leptonnumbers}
leave the Yukawa couplings invariant but they are broken \emph{softly}\/
by the Majorana mass terms of the right-handed neutrinos,
given by
\be
\label{majorana}
\mathcal{L}_\mathrm{Maj} =
- \frac{1}{2} \left( \begin{array}{ccc}
  \overline{\nu_{e R}}, & \overline{\nu_{\mu R}}, & \overline{\nu_{\tau R}}
\end{array} \right)
M_R C
\left( \begin{array}{c}
  \overline{\nu_{e R}}^T \\ \overline{\nu_{\mu R}}^T \\
  \overline{\nu_{\tau R}}^T
\end{array} \right)
+ \mathrm{H.c.}
\ee
In equation~\eqref{majorana},
$C$ is the charge-conjugation matrix in Dirac space;
the $3 \times 3$ flavour-space matrix $M_R$ is symmetric.
The seesaw mechanism produces an effective light-neutrino mass matrix
$M = - M_D M_R^{-1} M_D$,
\textit{i.e.}
\be
\label{MMM}
M_{\psi {\psi^\prime}} = - D_\psi D_{\psi^\prime}
\left( M_R^{-1} \right)_{\psi \psi^\prime}, \quad \forall\, \psi, \psi^\prime
\in \left\{ e,\, \mu,\, \tau \right\}.
\ee
Note that,
since $M_D$ and $M_\ell$ are diagonal,
the matrix $M_R$ is the sole origin
of lepton mixing in our
models~\cite{maximal}.\footnote{In the study of leptogenesis one uses
  a basis for the mass matrices where $M_\ell$ and $M_R$ are diagonal
  but $M_D$ is not.
  In our models $M_\ell$ and $M_D$ are diagonal but $M_R$ is not.
  If in our models $M_R$ is diagonalized as
  $V^T M_R V = \mathrm{diag} \left( M_1, M_2, M_3 \right)$,
  where $V$ is a $3 \times 3$ unitary matrix
  and $M_{1,2,3}$ are non-negative real,
  then $M_D^\prime = V^T M_D$
  in the basis appropriate for the computation of leptogenesis.
  The Hermitian matrix
  relevant for leptogenesis is $R \equiv M_D^\prime {M_D^\prime}^\dagger
  = V^T M_D M_D^\dagger V^\ast$,
  which is non-diagonal.
  Thus,
  leptogenesis is in principle viable in our models.
  See ref.~\cite{lepto} for details.}
The symmetric matrix $M$ is diagonalized as
\be
U^T M U = \mathrm{diag} \left( m_1,\, m_2,\, m_3 \right),
\ee
where the $m_k$ are the (non-negative real) light-neutrino masses
and $U$ is the lepton mixing matrix,
for which we use the standard parameterization~\cite{RPP}
\ba
U &=& \left( \begin{array}{ccc}
  c_{12} c_{13} & s_{12} c_{13} & s_{13} e^{- i \delta} \\
  - s_{12} c_{23} - c_{12} s_{23} s_{13} e^{i \delta} &
  c_{12} c_{23} - s_{12} s_{23} s_{13} e^{i \delta} &
  s_{23} c_{13} \\
  s_{12} s_{23} - c_{12} c_{23} s_{13} e^{i \delta} &
  - c_{12} s_{23} - s_{12} c_{23}s_{13} e^{i \delta} &
  c_{23} c_{13}
\end{array} \right)
\nonumber \\*[1mm] & &
\times\, \mathrm{diag} \left( 1,\ e^{i \alpha_{21} / 2},\
e^{i \alpha_{31} / 2} \right),
\label{uparam}
\ea
where $c_{ij} = \cos{\theta_{ij}}$ and $s_{ij} = \sin{\theta_{ij}}$
for $ij = 12, 23, 13$.
The phase $\delta$ is the Dirac phase;
$\alpha_{21}$ and $\alpha_{31}$ are the Majorana phases.

The purpose of our models is to make predictions for the matrix $U$.
There are in the literature many predictive models
for $U$;\footnote{See the reviews~\cite{mixing};
  the original papers are in the bibliographies of those reviews.}
the models in this paper are original in that
they are \emph{well-defined renormalizable models}\/
that produce \emph{predictions for the neutrino mass matrix $M$
  in terms of charged-lepton mass ratios}.
Since the mass ratios $m_\mu / m_\tau$,
$m_e / m_\mu$,
and $m_e / m_\tau$ are very small,
the predictions of our models are hardly distinguishable in practice
from the cases with `texture zeroes'
in the neutrino mass matrix~\cite{marfatia}.

In section~\ref{sec:models} we expound the construction of the models
and classify the various models that our class of models encompasses.
Section~\ref{sec:specific} focusses on a specific model
with remarkable predictions:
almost-maximal $\delta$
and almost-maximal neutrinoless double-beta decay.
Section~\ref{sec:potential} discusses a scalar potential for our models
and the way in which that potential is able to reproduce the Higgs particle
discovered at the LHC.
Section~\ref{sec:conclusions} contains our main conclusions.
Appendix~\ref{app:unitarity} deals on the derivation of the unitarity bounds
on the coupling constants of the scalar potential.
In appendix~\ref{stab} we compute the expectation value of the scalar potential
in the various stability points of that potential.

\section{Models} \label{sec:models}

\subsection{Construction of the models}

Our class of models may be divided into four subclasses:
\begin{enumerate}
\item Models with Yukawa Lagrangian
  \ba
  \mathcal{L}_1 &=&
  - y_1 \overline{D_{L \alpha}} \nu_{R \alpha} \tilde \phi_1
  - y_2 \overline{D_{L \alpha}} \alpha_R \phi_1
  \no & &
  - y_3 \left( \overline{D_{L \beta}} \nu_{R \beta} \tilde \phi_2
  + \overline{D_{L \gamma}} \nu_{R \gamma} \tilde \phi_3 \right)
  \no & &
  - y_4 \left( \overline{D_{L \beta}} \beta_R \phi_2
  + \overline{D_{L \gamma}} \gamma_R \phi_3 \right) + \mathrm{H.c.},
  \label{1}
  \ea
  where $\left( \alpha,\, \beta,\, \gamma \right)$ is a permutation
  of $\left( e,\, \mu,\, \tau \right)$
  and $y_{1,2,3,4}$ are Yukawa coupling constants,
  which are in general complex.
\item Models with Yukawa Lagrangian
  \ba
  \mathcal{L}_2 &=&
  - y_1 \overline{D_{L \alpha}} \nu_{R \alpha} \tilde \phi_1
  - y_2 \overline{D_{L \alpha}} \alpha_R \phi_1
  \no & &
  - y_3 \left( \overline{D_{L \beta}} \nu_{R \beta} \tilde \phi_2
  + \overline{D_{L \gamma}} \nu_{R \gamma} \tilde \phi_3 \right)
  \no & &
  - y_4 \left( \overline{D_{L \beta}} \beta_R \phi_3
  + \overline{D_{L \gamma}} \gamma_R \phi_2 \right) + \mathrm{H.c.}
  \label{2}
  \ea
  Note that $\mathcal{L}_1$ and $\mathcal{L}_2$
  differ only in their last lines.
\item Models with Yukawa Lagrangian
  \ba
  \mathcal{L}_3 &=&
  - y_1 \overline{D_{L \alpha}} \nu_{R \alpha} \tilde \phi_1
  - y_2 \overline{D_{L \alpha}} \alpha_R \phi_1
  \no & &
  - y_3 \overline{D_{L \beta}} \nu_{R \beta} \tilde \phi_2
  - y_3^\ast \overline{D_{L \gamma}} \nu_{R \gamma} \tilde \phi_3
  \no & &
  - y_4 \overline{D_{L \beta}} \beta_R \phi_2
  - y_4^\ast \overline{D_{L \gamma}} \gamma_R \phi_3 + \mathrm{H.c.},
  \label{3}
  \ea
  where $y_1$ and $y_2$ are real while $y_3$ and $y_4$ are in general complex.
\item Models with Yukawa Lagrangian
  \ba
  \mathcal{L}_4 &=&
  - y_1 \overline{D_{L \alpha}} \nu_{R \alpha} \tilde \phi_1
  - y_2 \overline{D_{L \alpha}} \alpha_R \phi_1
  \no & &
  - y_3 \overline{D_{L \beta}} \nu_{R \beta} \tilde \phi_2
  - y_3^\ast \overline{D_{L \gamma}} \nu_{R \gamma} \tilde \phi_3
  \no & &
  - y_4 \overline{D_{L \beta}} \beta_R \phi_3
  - y_4^\ast \overline{D_{L \gamma}} \gamma_R \phi_2 + \mathrm{H.c.},
  \label{4}
  \ea
  where once again $y_1$ and $y_2$ are real.
  The Lagrangians~\eqref{3} and~\eqref{4} differ in their last lines.
\end{enumerate}

It is clear that $\mathcal{L}_{1,2,3,4}$
enjoy the family-lepton-number symmetries~\eqref{leptonnumbers}.
The Lagrangians~\eqref{1} and~\eqref{2} further enjoy the interchange symmetry
\be
\label{interchange}
\phi_2 \leftrightarrow \phi_3, \quad
D_{L\beta} \leftrightarrow D_{L\gamma}, \quad
\beta_R \leftrightarrow \gamma_R, \quad
\nu_{R \beta} \leftrightarrow \nu_{R \gamma}.
\ee
The Lagrangians~\eqref{3} and~\eqref{4} are invariant under
the $CP$ symmetry
\be \label{CP}
\!\!
\begin{array}{l}
\phi_1 \left( x \right) \to \phi_1^\ast \left( \bar x \right), \quad
\phi_2 \left( x \right) \to \phi_3^\ast \left( \bar x \right), \quad
\phi_3 \left( x \right) \to \phi_2^\ast \left( \bar x \right), \quad
\\*[1mm]
\alpha_R \left( x \right) \to
K\, \overline{\alpha_R}^T \! \left( \bar x \right), \quad
\beta_R \left( x \right) \to
K\, \overline{\gamma_R}^T \! \left( \bar x \right), \quad
\gamma_R \left( x \right) \to
K\, \overline{\beta_R}^T \! \left( \bar x \right), \\*[1mm]
\nu_{R \alpha} \left( x \right) \to
K\, \overline{\nu_{R \alpha}}^T \! \left( \bar x \right), \quad
\nu_{R \beta} \left( x \right) \to
K\, \overline{\nu_{R \gamma}}^T \! \left( \bar x \right), \quad
\nu_{R \gamma} \left( x \right) \to
K\, \overline{\nu_{R \beta}}^T \! \left( \bar x \right), \\*[1mm]
D_{L \alpha} \left( x \right) \to
K \overline{D_{L \alpha}}^T \! \left( \bar x \right), \quad
D_{L \beta} \left( x \right) \to
K \overline{D_{L \gamma}}^T \! \left( \bar x \right), \quad
D_{L \gamma} \left( x \right) \to
K \overline{D_{L \beta}}^T \! \left( \bar x \right),
\end{array}
\ee
where $x \equiv \left( t,\, \vec r \right)$ and
$\bar x \equiv \left( t,\, - \vec r \right)$;
$K \equiv i \gamma_0 C$ is the $CP$-transformation matrix in Dirac space.
Moreover,
in the last line of transformation~\eqref{CP},
\be
\overline{D_{L \psi}}^T \equiv
\left( \begin{array}{c}
  \overline{\nu_{L \psi}}^T \\ \overline{\psi_L}^T
\end{array} \! \right).
\ee
The $CP$ transformation~\eqref{CP}
interchanges the lepton flavours $\beta$ and $\gamma$.

The Lagrangians~\eqref{1}--\eqref{4} necessitate
additional symmetries to guarantee that each scalar doublet
only couples to the desired lepton flavour.
There is a large arbitrariness in the choice of the additional symmetries.
In this paper we choose them to be
\be
\label{sym1}
\mathbbm{Z}_2^{(1)}: \quad \phi_1 \to - \phi_1, \quad
D_{L \alpha} \to - D_{L \alpha},
\ee
for all four Lagrangians~\eqref{1}--\eqref{4};
and either
\be
\label{sym2}
\begin{array}{l}
  \mathbbm{Z}_2^{(2)}: \quad \phi_2 \to - \phi_2, \quad
  \beta_R \to - \beta_R, \quad
  \nu_{R \beta} \to - \nu_{R \beta},
  \\*[1mm]
  \mathbbm{Z}_2^{(3)}: \quad \phi_3 \to - \phi_3, \quad
  \gamma_R \to - \gamma_R, \quad
  \nu_{R \gamma} \to - \nu_{R \gamma},
\end{array}
\ee
for Lagrangians~\eqref{1} and~\eqref{3},
or else
\be
\label{sym3}
\begin{array}{l}
  \mathbbm{Z}_2^{(4)}: \quad \phi_2 \to - \phi_2, \quad
  \gamma_R \to - \gamma_R, \quad
  \nu_{R \beta} \to - \nu_{R \beta},
  \\*[1mm]
  \mathbbm{Z}_2^{(5)}: \quad \phi_3 \to - \phi_3, \quad
  \beta_R \to - \beta_R, \quad
  \nu_{R \gamma} \to - \nu_{R \gamma},
\end{array}
\ee
for Lagrangians~\eqref{2} and~\eqref{4}.
The transformations~\eqref{sym1} and either~\eqref{sym2} or~\eqref{sym3}
form a $\mathbbm{Z}_2 \times \mathbbm{Z}_2 \times \mathbbm{Z}_2$ symmetry.

Let $v_k$ denote the vacuum expectation value (VEV) of $\phi_k^0$.
Then,
from $\mathcal{L}_1$,
\be
\label{111}
\begin{array}{lclcl}
\left( M_\ell \right)_{\alpha \alpha} \equiv \ell_\alpha = y_2 v_1, & &
\left( M_\ell \right)_{\beta \beta} \equiv \ell_\beta = y_4 v_2, & &
\left( M_\ell \right)_{\gamma \gamma} \equiv \ell_\gamma = y_4 v_3,
\\*[1mm]
\left( M_D \right)_{\alpha \alpha} \equiv D_\alpha = y_1^\ast v_1, & &
\left( M_D \right)_{\beta \beta} \equiv D_\beta = y_3^\ast v_2, & &
\left( M_D \right)_{\gamma \gamma} \equiv D_\gamma = y_3^\ast v_3
\end{array}
\ee
for model~1.
From the Yukawa Lagrangian~\eqref{2},
\be
\label{222}
\begin{array}{lclcl}
\ell_\alpha = y_2 v_1, & & \ell_\beta = y_4 v_3, & & \ell_\gamma= y_4 v_2,
\\*[1mm]
D_\alpha = y_1^\ast v_1, & & D_\beta = y_3^\ast v_2, & & D_\gamma = y_3^\ast v_3
\end{array}
\ee
for model~2.
From $\mathcal{L}_3$,
\be
\label{333}
\begin{array}{lclcl}
\ell_\alpha = y_2 v_1, & & \ell_\beta = y_4 v_2, & & \ell_\gamma = y_4^\ast v_3,
\\*[1mm]
D_\alpha = y_1^\ast v_1, & & D_\beta = y_3^\ast v_2, & & D_\gamma = y_3 v_3
\end{array}
\ee
for model~3.
From the Yukawa Lagrangian~\eqref{4},
\be
\label{444}
\begin{array}{lclcl}
\ell_\alpha = y_2 v_1, & & \ell_\beta = y_4 v_3, & & \ell_\gamma = y_4^\ast v_2,
\\*[1mm]
D_\alpha = y_1^\ast v_1, & & D_\beta = y_3^\ast v_2, & & D_\gamma = y_3 v_3
\end{array}
\ee
for model~4.

We next consider the right-handed-neutrino Majorana mass terms.
They softly break the lepton-number symmetries~\eqref{leptonnumbers}
and also the additional symmetries~\eqref{sym2} or~\eqref{sym3}.
We \emph{assume}\/ that they do \emph{not}\/ break
either the interchange symmetry~\eqref{interchange} of models~1 and~2
or the $CP$ symmetry~\eqref{CP} of models~3 and~4.
This means that,
in models~1 and~2,
\be
\label{buyfk}
\left( M_R \right)_{\beta \beta} = \left( M_R \right)_{\gamma \gamma}, \quad
\left( M_R \right)_{\alpha \beta} = \left( M_R \right)_{\alpha \gamma}.
\ee
Clearly,
the symmetry~\eqref{buyfk} for the matrix $M_R$
is also valid for the matrix $M_R^{-1}$.
Therefore,
from equation~\eqref{MMM},
\be
\frac{M_{\beta \beta}}{M_{\gamma \gamma}} = \left( \frac{D_\beta}{D_\gamma} \right)^2,
\quad
\frac{M_{\alpha \beta}}{M_{\alpha \gamma}} = \frac{D_\beta}{D_\gamma}
\ee
for models~1 and~2.
This means that the rephasing-invariant phase
\be
\label{arg12}
\arg{\left[ M_{\gamma \gamma} \left( M_{\alpha \beta} \right)^2
    M^\ast_{\beta \beta} \left( M^\ast_{\alpha \gamma} \right)^2 \right]} = 0
\ee
in models~1 and~2.
Additionally,
from equations~\eqref{111} and~\eqref{malpha},
\be
\label{mod1}
\left| \frac{M_{\beta \beta}}{M_{\gamma \gamma}} \right| = \frac{m_\beta^2}{m_\gamma^2},
\quad
\left| \frac{M_{\alpha \beta}}{M_{\alpha \gamma}} \right| = \frac{m_\beta}{m_\gamma}
\ee
for model~1;
while,
from equations~\eqref{222} and~\eqref{malpha},
\be
\label{mod2}
\left| \frac{M_{\beta \beta}}{M_{\gamma \gamma}} \right| = \frac{m_\gamma^2}{m_\beta^2},
\quad
\left| \frac{M_{\alpha \beta}}{M_{\alpha \gamma}} \right| = \frac{m_\gamma}{m_\beta}
\ee
for model~2.

We conclude that \emph{model~1 makes three predictions
for the effective light-neutrino mass matrix $M$:
equations~\eqref{arg12} and~\eqref{mod1}.
Model~2 also makes three predictions:
equations~\eqref{arg12} and~\eqref{mod2}.}

In models~3 and~4,
we assume that the $CP$ symmetry~\eqref{CP} is not broken
by the Majorana mass terms of the $\nu_R$.
This means that
\be
\label{buyfk2}
\begin{array}{lcl}
\left( M_R \right)_{\beta \beta} = \left( M_R^\ast \right)_{\gamma \gamma}, & &
\left( M_R \right)_{\alpha \beta} = \left( M_R^\ast \right)_{\alpha \gamma},
\\*[1mm]
\left( M_R \right)_{\alpha \alpha} = \left( M_R^\ast \right)_{\alpha \alpha}, & &
\left( M_R \right)_{\beta \gamma} = \left( M_R^\ast \right)_{\beta \gamma}
\end{array}
\ee
in those models.
Equations~\eqref{buyfk2} are valid for $M_R^{-1}$ as weall as for $M_R$,
hence
\be
\label{bubiho}
\frac{M_{\beta \beta}}{M_{\gamma \gamma}^\ast}
= \frac{D_\beta D_\beta}{D_\gamma^\ast D_\gamma^\ast},
\quad
\frac{M_{\alpha \beta}}{M_{\alpha \gamma}^\ast}
= \frac{D_\alpha D_\beta}{D_\alpha^\ast D_\gamma^\ast}
\quad
\frac{M_{\alpha \alpha}}{M_{\alpha \alpha}^\ast}
= \frac{D_\alpha D_\alpha}{D_\alpha^\ast D_\alpha^\ast},
\quad
\frac{M_{\beta \gamma}}{M_{\beta \gamma}^\ast}
= \frac{D_\beta D_\gamma}{D_\beta^\ast D_\gamma^\ast}
\ee
for models~3 and~4.
Equations~\eqref{bubiho} imply the following
rephasing-invariant conditions on the matrix $M$:
\bs
\label{args}
\ba
\arg{\left[ M^\ast_{\beta \beta} M^\ast_{\gamma \gamma} \left( M_{\beta \gamma} \right)^2
    \right]} &=& 0,
\\
\arg{\left( M^\ast_{\alpha \alpha} M^\ast_{\beta \gamma}
    M_{\alpha \beta} M_{\alpha \gamma}
    \right)} &=& 0.
\ea
\es
Moreover,
from equations~\eqref{malpha} and~\eqref{333}
one derives equation~\eqref{mod1},
which is thus also valid for model~3;
from equations~\eqref{malpha} and~\eqref{444}
one derives equation~\eqref{mod2},
which thus applies to model~4.
We conclude that \emph{model~3 makes four predictions for $M$:
equations~\eqref{args} and~\eqref{mod1}.
Model~4 also makes four predictions:
equations~\eqref{args} and~\eqref{mod2}.}

\subsection{Classification of the models}

Our class of models encompasses twelve models,
depending on whether one uses model~1,
2,
3,
or~4 and depending on whether the flavour $\alpha$ is taken to be~$e$,
$\mu$,
or~$\tau$.
(The flavours $\beta$ and $\gamma$ are treated symetrically in the models.)

There is a distinction between the models
with interchange symmetry~\eqref{interchange}
and the models with $CP$ symmetry~\eqref{CP}:
the former lead to only one constraint~\eqref{arg12}
on the phases of the matrix elements of $M$,
while the latter lead to the two constraints~\eqref{args}.
The $CP$ symmetry~\eqref{CP} is more powerful
than the interchange symmetry~\eqref{interchange}.

However,
in practice the distinction between equation~\eqref{arg12}
and equations~\eqref{args} is not very significant,
because the charged-lepton mass ratios are so small
that they force some $M$-matrix elements to be very close to zero,
hence their phases do not matter much.
We see from equations~\eqref{mod1} and~\eqref{mod2} that our twelve models
may be classified in six types:
\begin{enumerate}
  \renewcommand{\labelenumi}{\roman{enumi}.}
\item Models that predict
  \be
  \left| \frac{M_{ee}}{M_{\mu \mu}} \right| = \frac{m_e^2}{m_\mu^2},
  \quad
  \left| \frac{M_{e\tau}}{M_{\mu \tau}} \right| = \frac{m_e}{m_\mu}.
  \ee
  Since $m_e \ll m_\mu$,
  in these models one is close to the situation $M_{ee} = M_{e\tau} = 0$,
  which is case A$_2$ of ref.~\cite{marfatia}.
\item Models that predict
  \be
  \left| \frac{M_{ee}}{M_{\mu\mu}} \right| = \frac{m_\mu^2}{m_e^2},
  \quad
  \left| \frac{M_{e\tau}}{M_{\mu\tau}} \right| = \frac{m_\mu}{m_e}.
  \ee
  Since $m_e \ll m_\mu$ these models predict $M_{\mu\mu} \approx 0$
  and $M_{\mu\tau} \approx 0$.
  According to ref.~\cite{marfatia},
  $M_{\mu\mu} = M_{\mu\tau} = 0$ is phenomenologically excluded.
\item Models that predict
  \be
  \left| \frac{M_{ee}}{M_{\tau\tau}} \right| = \frac{m_e^2}{m_\tau^2},
  \quad
  \left| \frac{M_{e\mu}}{M_{\mu \tau}} \right| = \frac{m_e}{m_\tau}.
  \ee
  Since $m_e \ll m_\tau$ these models predict $M_{ee} \approx 0$
  and $M_{e\mu} \approx 0$.
  They are therefore close to case A$_1$ of ref.~\cite{marfatia}.
\item Models that predict
  \be
  \left| \frac{M_{ee}}{M_{\tau\tau}} \right| = \frac{m_\tau^2}{m_e^2},
  \quad
  \left| \frac{M_{e\mu}}{M_{\mu \tau}} \right| = \frac{m_\tau}{m_e}.
  \ee
  This leads to,
  approximately,
  $M_{\tau\tau} = M_{\mu\tau} = 0$,
  which is phenomenologically excluded.
\item Models that predict
  \be
  \left| \frac{M_{\mu\mu}}{M_{\tau\tau}} \right| = \frac{m_\mu^2}{m_\tau^2},
  \quad
  \left| \frac{M_{e\mu}}{M_{e\tau}} \right| = \frac{m_\mu}{m_\tau}.
  \ee
  Since $m_\mu \ll m_\tau$ these models predict $M_{\mu\mu} \approx 0$
  and $M_{e\mu} \approx 0$.
  They are therefore close to case B$_3$ of ref.~\cite{marfatia}.
\item Models that predict
  \be
  \left| \frac{M_{\mu\mu}}{M_{\tau\tau}} \right| = \frac{m_\tau^2}{m_\mu^2},
  \quad
  \left| \frac{M_{e\mu}}{M_{e\tau}} \right| = \frac{m_\tau}{m_\mu}.
  \ee
  This leads to $M_{\tau\tau} \approx 0$ and $M_{e\tau} \approx 0$,
  corresponding to case B$_4$ of ref.~\cite{marfatia}.
\end{enumerate}

We thus find that,
out of our twelve models,
four should be phenomenologically excluded.
The other eight are viable;
two of them approximately coincide in their predictions with case A$_1$
of ref.~\cite{marfatia},
two other with case A$_2$,
two more with case B$_3$,
and the last two with case B$_4$.

We have made numerical simulations of all our models and they very much
vindicate the above conclusions.
We do not feel it worth presenting those numerical simulations in detail here.
In the next section we focus solely on one model
that in our opinion yields particularly interesting results.

\section{A specific model} \label{sec:specific}

In this section we deal on one of our models,
which
predicts
\bs
\label{predi}
\ba
\left| \frac{M_{\mu\mu}}{M_{\tau\tau}} \right| &=& \frac{m_\mu^2}{m_\tau^2},
\label{11} \\
\left| \frac{M_{e\mu}}{M_{e\tau}} \right| &=& \frac{m_\mu}{m_\tau},
\label{22} \\
\arg{\left[ M_{\tau\tau} \left( M_{e\mu} \right)^2
    M^\ast_{\mu\mu} \left( M^\ast_{e\tau} \right)^2 \right]} &=& 0.
\label{33}
\ea
\es
Equations~\eqref{predi} are three predictions.
This is not much;
for instance,
each of the cases with two texture zeroes of ref.~\cite{marfatia}
has four predictions,
and there are models with as many as six predictions for $M$.
So,
one might think that the predictions~\eqref{predi}
are of little practical consequence.
This is not so,
however.

We use $M = U^\ast\, \mathrm{diag} \left( m_1,\, m_2,\, m_3 \right)
U^\dagger$ and the parameterization of $U$ in equation~\eqref{uparam}.
We also use the experimental $3\sigma$ bounds~\cite{tortola}
\bs
\label{gene}
\ba
& & 7.05 \le \frac{m_2^2 - m_1^2}{10^{-5}\, \mathrm{eV}^2} \le 8.14, \\*[1mm]
& & 0.273 \le s_{12}^2 \le 0.379, \\
& & 0.0189 \le s_{13}^2 \le 0.0239,
\ea
\es
and either
\bs
\label{norm}
\ba
& & 2.43 \le \frac{m_3^2 - m_1^2}{10^{-3}\, \mathrm{eV}^2} \le 2.67, \\*[1mm]
& & 0.384 \le s_{23}^2 \le 0.635
\ea
\es
for a normal ordering of the neutrino masses,
or
\bs
\label{inve}
\ba
& & 2.37 \le \frac{m_1^2 - m_3^2}{10^{-3}\, \mathrm{eV}^2} \le 2.61, \\*[1mm]
& & 0.388 \le s_{23}^2 \le 0.638
\ea
\es
for the inverted ordering of the neutrino masses.
The phases $\delta$,
$\alpha_{21}$,
and $\alpha_{31}$ are unknown,
just as the overall scale of the neutrino masses;
we represent the latter through $m_\mathrm{sum} \equiv m_1 + m_2 + m_3$.
Strong cosmological arguments suggest that $m_\mathrm{sum} \le 0.25\,$eV
at 95\%~confidence level~\cite{Planck}.\footnote{A recent paper~\cite{new}
  claims that $m_\mathrm{sum} = 0.11 \pm 0.03\,$eV.}

A quantity of especial importance is
\be
m_{\beta\beta} \equiv \left| M_{ee} \right| =
\left| m_1 c_{12}^2 c_{13}^2 + m_2 s_{12}^2 c_{13}^2 e^{i \alpha_{21}}
+ m_3 s_{13}^2 e^{i \left( \alpha_{31} - 2 \delta \right)} \right|.
\ee
This quantity is relevant for neutrinoless double-beta decay,
which should proceed with a rate approximately proportional
to $m_{\beta\beta}^2$.
It is clear that $m_{\beta\beta}$ becomes maximal
when
\be
\label{alphas}
\alpha_{21} = 0, \quad \alpha_{31} = 2 \delta,
\ee
for whatever value of the phase $\delta$.

\begin{figure}
\begin{center}
\epsfig{file=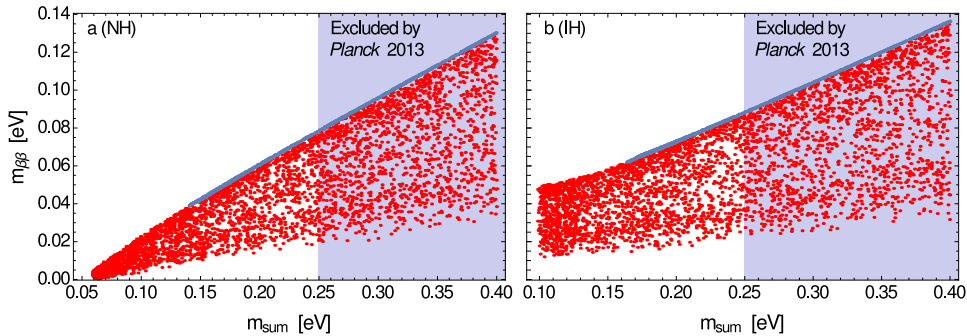,width=1.0\textwidth}
\end{center}
\caption{Scatter plots of $m_{\beta\beta}$ \textit{versus} $m_\mathrm{sum}$.
  Figure~\ref{fig10}a is for a normal ordering of the neutrino masses
  and figure~\ref{fig10}b is for an inverted ordering.
  The red points take into account only
  the experimental bounds~\eqref{gene}--\eqref{inve};
  the blue points arise from the constraints~\eqref{predi}.
  \label{fig10}}
\end{figure}

In figure~\ref{fig10}
we have plotted $m_{\beta\beta}$ as a function of $m_\mathrm{sum}$,
both when only the inequalities~\eqref{gene}
and either~\eqref{norm} or~\eqref{inve} hold,
and when furthermore the predictions~\eqref{predi} are enforced.
The information in that figure is clear:
the predictions~\eqref{predi} lead to almost maximal $m_{\beta\beta}$,
irrespective of the neutrino mass ordering.

\begin{figure}
\begin{center}
\epsfig{file=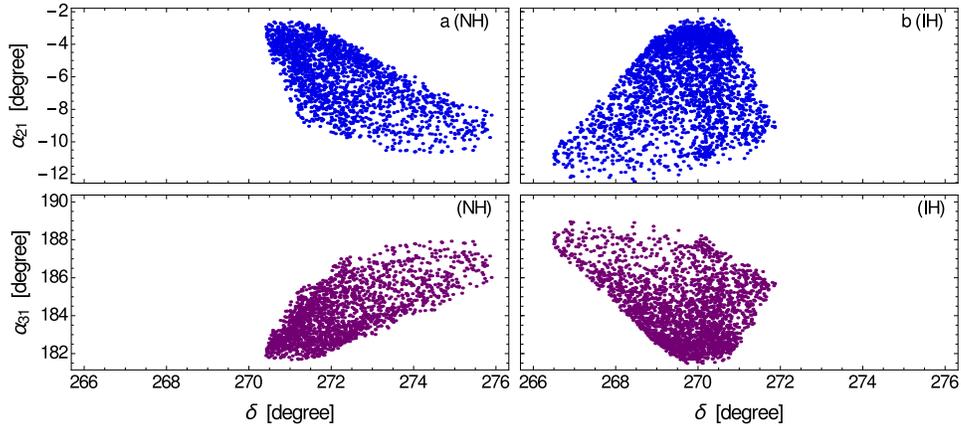,width=1.0\textwidth}
\end{center}
\caption{Scatter plots of the phases $\delta$,
  $\alpha_{21}$,
  and $\alpha_{31}$ following from the constraints~\eqref{predi}.
  Figure~\ref{fig11}a is for a normal ordering of the neutrino masses
  and figure~\ref{fig11}b is for an inverted ordering.
  \label{fig11}}
\end{figure}

This of course happens because equations~\eqref{alphas} hold.
In fig.~\ref{fig11}
one observes that this is indeed so and that,
moreover,
the predictions~\eqref{predi} lead to $\delta \approx 3 \pi / 2$.
Thus,
our model firmly predicts the three phases $\delta$,
$\alpha_{21}$,
and $\alpha_{31}$;
the phase $\delta$ is predicted to be very close to $1.5 \pi$,
and this agrees nicely with its $1\sigma$-preferred experimental
value~\cite{tortola}.

One moreover observes in figure~\ref{fig10} that our model
does not tolerate very low neutrino masses,
but goes well with almost-degenerate neutrinos:
$m_\mathrm{sum} \gtrsim 0.15\,$eV for both
the normal and inverted neutrino mass spectra.

\begin{figure}
\begin{center}
\epsfig{file=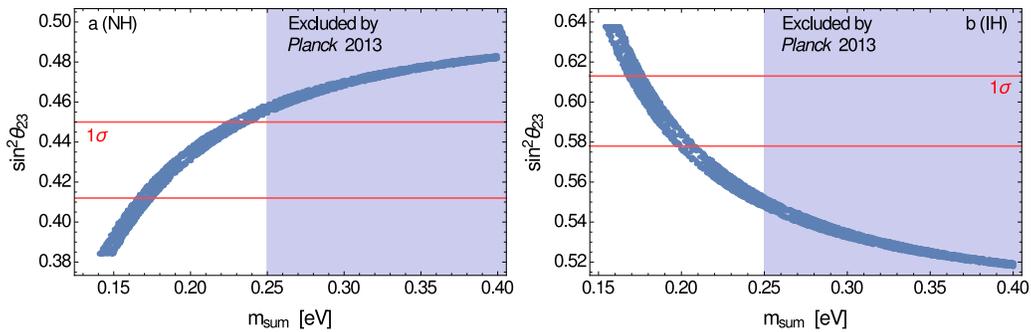,width=1.0\textwidth}
\end{center}
\caption{Scatter plots of $s_{23}^2$ \textit{versus} $m_\mathrm{sum}$
  following from the constraints~\eqref{predi}.
  Figure~\ref{fig12}a is for a normal ordering of the neutrino masses
  and figure~\ref{fig12}b is for an inverted ordering.
  \label{fig12}}
\end{figure}

This specific model does not just predict the Dirac and Majorana phases;
it moreover predicts the quadrant of the angle $\theta_{23}$
and a correlation between that angle and $m_\mathrm{sum}$.
That is observed in fig.~\ref{fig12}.
One sees that $\theta_{23}$ lies in the first quadrant
when the neutrino mass ordering is normal,
in the second quadrant when it is inverted.
One also sees that $\theta_{23}$ is correlated with $m_\mathrm{sum}$,
with $\theta_{23}$ becoming ever closer to $\pi / 4$
when $m_\mathrm{sum}$ grows.

Figures~\ref{fig10} and~\ref{fig12} are very similar
to analogous figures displayed in ref.~\cite{peinado}
for case~B$_3$ of ref.~\cite{marfatia}.
That case is \emph{defined} by
$M_{\mu\mu} = M_{e\mu} = 0$,\footnote{The paper of ref.~\cite{marfatia}
  contains various two-texture-zero cases,
  in particular case B$_3$ defined as $M_{\mu\mu} = M_{e\mu} = 0$.
  The cases are of course not full models.
  However,
  it was demonstrated in ref.~\cite{we} that
  \emph{any}\/ texture-zero mass matrix may result from a renormalizable model.}
which of course means four predictions for $M$
(because both the moduli and phases
of $M_{\mu\mu}$ and $M_{e\mu}$ are relevant).
Our predictions~\eqref{predi} mean that our model
features both $\left| M_{\mu\mu} \right| \ll \left| M_{\tau\tau} \right|$
and $\left| M_{e\mu} \right| \ll \left| M_{e\tau} \right|$,
and this is an approximation to case B$_3$.
As a matter of fact,
we have explicitly checked that
the two conditions~\eqref{11} and~\eqref{22} by themselves alone
lead to almost the same allowed domains as in figures~\ref{fig10}--\ref{fig12},
and as in case B$_3$ of ref.~\cite{marfatia}.
The two conditions~\eqref{11} and~\eqref{22}
are in practice just as predictive as that case with four predictions.


\section{The scalar potential} \label{sec:potential}

\subsection{Assumptions}

In this section we investigate a way in which our class of models
with three Higgs doublets and various symmetries may
(i) be extended to the quark sector,
and (ii) produce scalar particles with masses and couplings
in agreement with the phenomenology.
The aim of our investigation is to demonstrate that this can be done;
we do not explore the full set of options.
Thus,
in this section we make \emph{additional assumptions}.
We stress that
the validity of the models expounded in section~\ref{sec:models}
is in general independent of the specific additional assumptions
that we shall utilize in this section.

Our main assumption is that \emph{there are no scalars
  besides the three Higgs doublets}\/
that have Yukawa couplings to the leptons.
Therefore,
\be
\label{v}
v \equiv \sqrt{\left| v_1 \right|^2 + \left| v_2 \right|^2
  + \left| v_3 \right|^2} = \frac{\sqrt{2} m_W}{g}
\approx 174\, \mathrm{GeV},
\ee
where $m_W = 80.1\, \mathrm{GeV}$ is the mass of the $W^\pm$ bosons
and $g$ is the gauge-$SU(2)$ coupling constant.

Our models have either an interchange symmetry~\eqref{interchange}
or the $CP$ symmetry~\eqref{CP}.
Those symmetries are unbroken
by the Majorana mass terms of the right-handed neutrinos,
which have mass dimension three.
Still,
those symmetries may be broken by the quadratic
(\textit{i.e.}\ mass dimension two)
terms of the scalar potential.
We shall assume that this does not happen,
\textit{i.e.}\ that \emph{either the interchange symmetry~\eqref{interchange}
  or the $CP$ symmetry~\eqref{CP} are conserved by the quadratic terms
  of the scalar potential}.
The potential is thus symmetric under either
$\phi_2 \leftrightarrow \phi_3$ or $\phi_2 \leftrightarrow \phi_3^\ast$.
In this paper we shall only consider the potential invariant
under $\phi_2 \leftrightarrow \phi_3$;\footnote{Our potential
  \label{domainwalls}
  is therefore invariant under a $\mathbbm{Z}_2$ symmetry.
  When that $\mathbbm{Z}_2$ symmetry is spontaneously broken,
  the vacuum is two-fold degenerate.
  There is a minimum-energy field configuration that
  interpolates between the two different vacua;
  this is called a domain wall.
  The non-observation of domain walls definitely is a problem
  for our potential.
  However,
  we recall the reader that our analysis only purports to display
  a particularly simple and illustrative case;
  we claim our potential neither to be realistic nor to be unique.
  The validity of the models expounded in section~\ref{sec:models}
  is independent of the specific scalar potential
  that we analyze in this section.}
in ref.~\cite{previous} the potential invariant
under $\phi_2 \leftrightarrow \phi_3^\ast$ has been studied.

Besides,
the models have additional symmetries~\eqref{sym1} together with
either~\eqref{sym2} or~\eqref{sym3}.\footnote{The additional symmetries
  are largely arbitrary---in the construction of the models
  we might have chosen different additional symmetries
  to the same practical effect,
  \textit{viz.}~preventing each scalar doublet from having Yukawa couplings
  to more than one lepton doublet.
  Each specific additional symmmetry alters the scalar potential
  in a different way.
  Thus,
  in a sense the specific additional symmetries~\eqref{sym1}--\eqref{sym3}
  constitute \emph{an assumption}\/ of this section.}
The symmetry~\eqref{sym1} does not involve the $\nu_{R\psi}$
and is therefore unbroken by $\mathcal{L}_\mathrm{Maj}$.
We shall assume that it is also unbroken by the scalar potential;
thus,
the potential is invariant under
$\mathbbm{Z}_2^{(1)}\!:\, \phi_1 \to - \phi_1$.\footnote{In ref.~\cite{previous}
  a potential with quadratic terms $\phi_1^\dagger \phi_2$,
  $\phi_1^\dagger \phi_3$,
  and their Hermitian conjugates has been analyzed.
  The fit in this section does not allow for those terms,
  which break the symmetry $\mathbbm{Z}_2^{(1)}$.
  See also footnote~\ref{domainwalls}.}
The symmetries~\eqref{sym2} or~\eqref{sym3},
which read $\mathbbm{Z}_2^{(2,4)}\!:\, \phi_2 \to - \phi_2$
and $\mathbbm{Z}_2^{(3,5)}\!:\, \phi_3 \to - \phi_3$
in the scalar sector,
are softly broken by $\mathcal{L}_\mathrm{Maj}$,
which is of dimension three;
therefore,
they must also be broken in the quadratic part of the potential.
The potential therefore is
\ba
V &=& \mu_1\, \phi_1^\dagger \phi_1
+ \mu_2 \left( \phi_2^\dagger \phi_2 + \phi_3^\dagger \phi_3 \right)
+ \mu_3 \left( \phi_2^\dagger \phi_3 + \phi_3^\dagger \phi_2 \right)
\no & &
+ \lambda_1 \left( \phi_1^\dagger \phi_1 \right)^2
+ \lambda_2 \left[ \left( \phi_2^\dagger \phi_2 \right)^2
  + \left( \phi_3^\dagger \phi_3 \right)^2 \right]
\no & &
+ \lambda_3\, \phi_1^\dagger \phi_1 \left( \phi_2^\dagger \phi_2
+ \phi_3^\dagger \phi_3 \right)
+ \lambda_4\, \phi_2^\dagger \phi_2\, \phi_3^\dagger \phi_3
\no & &
+ \lambda_5 \left( \phi_1^\dagger \phi_2\, \phi_2^\dagger \phi_1
+ \phi_1^\dagger \phi_3\, \phi_3^\dagger \phi_1 \right)
+ \lambda_6\, \phi_2^\dagger \phi_3\, \phi_3^\dagger \phi_2
\no & &
+ \lambda_7 \left[ \left( \phi_1^\dagger \phi_2 \right)^2
  + \left( \phi_1^\dagger \phi_3 \right)^2 \right]
+ \lambda_7^\ast \left[ \left( \phi_2^\dagger \phi_1 \right)^2
  + \left( \phi_3^\dagger \phi_1 \right)^2 \right]
\no & &
+ \lambda_8 \left[ \left( \phi_2^\dagger \phi_3 \right)^2
  + \left( \phi_3^\dagger \phi_2 \right)^2 \right].
\label{vv}
\ea
The parameters $\mu_3$ and $\lambda_8$ are real because of the symmetry
under $\phi_2 \leftrightarrow \phi_3$.
We use the freedom of rephasing $\phi_1$ to set $\lambda_7$ real too.

\subsection{The vacuum}

We assume as usual that the vacuum
does not break electromagnetic invariance,
\textit{i.e.}\ that the upper components of $\phi_{1,2,3}$ have zero VEV.

The potential~\eqref{vv} may,
at least for some values of its parameters,
produce stability points with non-trivial relative phases among the VEVs.
Those stability points are,
unfortunately,
hard to manipulate analytically.
We shall neglect them and
\emph{assume that the three VEVs
  $v_k \equiv \left\langle 0 \left| \phi_k^0 \right| 0 \right\rangle$
  are (relatively) real}.
The VEV of the potential is then
\ba
V_0 \equiv \left\langle 0 \left| V \right| 0 \right\rangle &=&
\mu_1 v_1^2 + \mu_2 \left( v_2^2 + v_3^2 \right) + 2 \mu_3 v_2 v_3
\no & &
+ \lambda_1 v_1^4 + \lambda_2 \left( v_2^4 + v_3^4 \right)
+ 2 l_3 v_1^2 \left( v_2^2 + v_3^2 \right)
+ 2 l_4 v_2^2 v_3^2, \hspace*{7mm}
\label{v0}
\ea
where
\bs
\label{lll}
\ba
l_3 &\equiv& \frac{\lambda_3 + \lambda_5}{2} + \lambda_7, \\
l_4 &\equiv& \frac{\lambda_4 + \lambda_6}{2} + \lambda_8.
\ea
\es
The equations for vacuum stability are
\bs
\label{vs}
\ba
0 = \frac{\partial V_0}{\partial v_1^2} &=&
\mu_1 + 2 \lambda_1 v_1^2 + 2 l_3 \left( v_2^2 + v_3^2 \right),
\label{ve} \\
0 = \frac{1}{2}\, \frac{\partial V_0}{\partial v_2} &=&
\mu_2 v_2 + \mu_3 v_3
+ 2 \lambda_2 v_2^3 + 2 l_3 v_1^2 v_2 + 2 l_4 v_2 v_3^2,
\label{vmu} \\
0 = \frac{1}{2}\, \frac{\partial V_0}{\partial v_3} &=&
\mu_2 v_3 + \mu_3 v_2
+ 2 \lambda_2 v_3^3 + 2 l_3 v_1^2 v_3 + 2 l_4 v_2^2 v_3.
\label{vtau}
\ea
\es

We want a vacuum state with $v_1 \neq 0$,
because in our models
one of the charged-lepton masses is proportional to $\left| v_1 \right|$.
We also want the vacuum to have $\left| v_2 \right| \neq \left| v_3 \right|$,
because in our models $r \equiv \left| v_2 / v_3 \right|$
is equal to a ratio of charged-lepton masses.
Fortunately,
equations~\eqref{vs} have a solution with $v_1 \neq 0$ and $v_2 \neq \pm v_3$:
\bs
\label{vugio}
\ba
\mu_1 &=& - 2 \lambda_1 v_1^2 - 2 l_3 \left( v_2^2 + v_3^2 \right), \\
\mu_2 &=& - 2 l_3 v_1^2 - 2 \lambda_2 \left( v_2^2 + v_3^2 \right), \\
\mu_3 &=& 2 \left( \lambda_2 - l_4 \right) v_2 v_3.
\ea
\es
Plugging equations~\eqref{vugio} into equation~\eqref{v0},
we obtain
\be
\label{vvv000}
V_0 = \frac{\mu_3^2}{2 \left( \lambda_2 - l_4 \right)}
+ \frac{\lambda_2 \mu_1^2 + \lambda_1 \mu_2^2 - 2 l_3 \mu_1 \mu_2}{4
  \left( l_3^2 - \lambda_1 \lambda_2 \right)}.
\ee

We parameterize
\bs
\label{vbeta}
\ba
v_1 &=& v \sin{\beta}, \\
v_2 &=& - \frac{v r \cos{\beta}}{\sqrt{1 + r^2}}, \\
v_3 &=& \frac{v \cos{\beta}}{\sqrt{1 + r^2}},
\ea
\es
and we use $r = m_\mu / m_\tau$
(the results for either $r = m_e / m_\mu$ or $r = m_e / m_\tau$
are not qualitatively different).
The angle $\beta$ will be taken to lie in the first quadrant.
In this way $v_1$ and $v_3$ are positive,
but this represents no lack of generality.
Only the relative sign of $v_2$ and $v_3$ matters,
and we have found out that the best results
are obtained when $v_2 v_3$ is negative.

\subsection{The scalar mass matrices}

We expand the neutral components of the doublets as
\be
\label{phialpha}
\phi_k^0 = v_k + \frac{\rho_k + i \eta_k}{\sqrt{2}},
\ee
where the fields $\rho_k$ and $\eta_k$ are real.
Subsuming the terms of the potential quadratic in the fields as
\bs
\ba
V_\mathrm{quadratic} &=&
\frac{1}{2} \left( \begin{array}{ccc} \eta_1, & \eta_2, & \eta_3
\end{array} \right) M_\eta \left( \begin{array}{c} \eta_1 \\ \eta_2 \\ \eta_3
\end{array} \right)
\\ & &
+ \frac{1}{2} \left( \begin{array}{ccc} \rho_1, & \rho_2, & \rho_3
\end{array} \right) M_\rho \left( \begin{array}{c} \rho_1 \\ \rho_2 \\ \rho_3
\end{array} \right)
\\ & &
+ \left( \begin{array}{ccc} \phi_1^-, & \phi_2^-, & \phi_3^-
\end{array} \right) M_\phi \left( \begin{array}{c} \phi_1^+ \\ \phi_2^+ \\
  \phi_3^+ \end{array} \right),
\ea
\es
we find,
by using equations~\eqref{vugio},
that
\bs
\label{mmm}
\ba
M_\eta &=& 4 \lambda_7 \left( \begin{array}{ccc}
  - v_2^2 - v_3^2 & v_1 v_2 & v_1 v_3 \\
  v_1 v_2 & - v_1^2 & 0 \\
  v_1 v_3 & 0 & - v_1^2
\end{array} \right)
\no & &
+ \left( 2 \lambda_2 - \lambda_4 - \lambda_6 + 2 \lambda_8 \right)
\left( \begin{array}{ccc}
  0 & 0 & 0 \\ 0 & - v_3^2 & v_2 v_3 \\ 0 & v_2 v_3 & - v_2^2
\end{array} \right),
\label{MI} \\
M_\phi &=& \left( \lambda_5 + 2 \lambda_7 \right) \left( \begin{array}{ccc}
  - v_2^2 - v_3^2 & v_1 v_2 & v_1 v_3 \\
  v_1 v_2 & - v_1^2 & 0 \\
  v_1 v_3 & 0 & - v_1^2
\end{array} \right)
\no & &
+ \left( 2 \lambda_2 - \lambda_4 \right) \left( \begin{array}{ccc}
  0 & 0 & 0 \\ 0 & - v_3^2 & v_2 v_3 \\ 0 & v_2 v_3 & - v_2^2
\end{array} \right),
\label{MC} \\
M_\rho &=& 4 \left( \begin{array}{ccc}
  \lambda_1 v_1^2 & l_3 v_1 v_2 & l_3 v_1 v_3 \\
  l_3 v_1 v_2 & \lambda_2 v_2^2 & 0 \\
  l_3 v_1 v_3 & 0 & \lambda_2 v_3^2
\end{array} \right)
\no & &
+ 2 \left( \begin{array}{ccc}
  0 & 0 & 0 \\
  0 & \left( l_4 - \lambda_2 \right) v_3^2 &
  \left( l_4 + \lambda_2 \right) v_2 v_3 \\
  0 & \left( l_4 + \lambda_2 \right) v_2 v_3 &
  \left( l_4 - \lambda_2 \right) v_2^2
\end{array} \right).
\label{MR}
\ea
\es

In general,
the matrices $M_\eta$ and $M_\phi$ must have an eigenvector
$\left( v_1,\, v_2,\, v_3 \right)$ with eigenvalue zero,
corresponding to the Goldstone bosons,
hence they must be of form
\ba
M_{\eta, \phi} &=&
a_{\eta,\phi} \left( \begin{array}{ccc}
  v_2^2 & - v_1 v_2 & 0 \\ - v_1 v_2 & v_1^2 & 0 \\ 0 & 0 & 0
\end{array} \right)
+ b_{\eta,\phi} \left( \begin{array}{ccc}
  v_3^2 & 0 & - v_1 v_3 \\ 0 & 0 & 0 \\ - v_1 v_3 & 0 & v_1^2
\end{array} \right)
\no & &
+ c_{\eta,\phi} \left( \begin{array}{ccc}
   0 & 0 & 0 \\ 0 & v_3^2 & - v_2 v_3 \\ 0 & - v_2 v_3 & v_2^2
\end{array} \right).
\ea
In our specific case,
due to the $\phi_2 \leftrightarrow \phi_3$ symmetry of $V$,
the coefficients $a_{\eta,\phi} = b_{\eta,\phi}$.
This has the important consequence that
both $M_\eta$ and $M_\phi$ are diagonalized by the orthogonal matrix
\be
O_v = \left( \begin{array}{ccc}
  v_1 / v & 0 & - v_{23} / v \\
  v_2 / v & v_3 / v_{23} & v_1 v_2 / \left( v v_{23} \right) \\
  v_3 / v & - v_2 / v_{23} & v_1 v_3 / \left( v v_{23} \right)
\end{array} \right),
\ee
where $v_{23} \equiv \sqrt{v_2^2 + v_3^2} = v \cos{\beta}$.
We find that
\bs
\ba
M_\eta^\prime \equiv O_v^T M_\eta O_v &=&
\mathrm{diag} \left( 0,\ m_{A_2}^2,\ m_{A_3}^2 \right),
\\
M_\phi^\prime \equiv O_v^T M_\phi O_v &=&
\mathrm{diag} \left( 0,\ m_{\varphi_2}^2,\ m_{\varphi_3}^2 \right),
\ea
\es
where
\bs
\label{themass}
\ba
m_{A_2}^2 &=& - 4 \lambda_7 v_1^2 -
\left( 2 \lambda_2 - \lambda_4 - \lambda_6 + 2 \lambda_8 \right) v_{23}^2,
\\
m_{A_3}^2 &=& - 4 \lambda_7 v^2,
\\
m_{\varphi_2}^2 &=& - \left( \lambda_5 + 2 \lambda_7 \right) v_1^2 -
\left( 2 \lambda_2 - \lambda_4 \right) v_{23}^2,
\\
m_{\varphi_3}^2 &=& - \left( \lambda_5 + 2 \lambda_7 \right) v^2.
\ea
\es

We diagonalize $M_\rho$ as
\be
\label{diag}
O_\rho^T M_\rho\, O_\rho
= \mathrm{diag} \left( m_{H_1}^2,\, m_{H_2}^2,\, m_{H_3}^2 \right),
\ee
where $O_\rho$ is a real,
orthogonal matrix.
We order its columns in such a way that $m_{H_1}^2 \le m_{H_2}^2 \le m_{H_3}^2$.
The fields $H_k = \sum_{k^\prime = 1}^3 \rho_{k^\prime} \left( O_\rho \right)_{k^\prime k}$
are physical scalars with mass $m_{H_k}$.

\subsection{The oblique parameter $T$} \label{t}

Defining
\be
F \left( x, y \right) \equiv \left\{ \begin{array}{lcl}
  \displaystyle{\frac{x + y}{2} - \frac{x y}{x - y}\, \ln{\frac{x}{y}}}
&\Leftarrow& x \neq y, \\*[3mm] 0 &\Leftarrow& x = y, \end{array} \right. 
\ee
the oblique parameter $T$ is~\cite{Grimus:2007if}
\ba
T &=& \frac{1}{16 \pi s_w^2 m_W^2} \left\{
\sum_{k=2}^3 F \left( m_{\varphi_k}^2, m_{A_k}^2 \right)
\right. \no & &
+ \sum_{k=2}^3 \sum_{k^\prime=1}^3
\left| \left( O_v^T O_\rho \right)_{kk^\prime} \right|^2
\left[ F \left( m_{\varphi_k}^2, m_{H_{k^\prime}}^2 \right)
  - F \left( m_{A_k}^2, m_{H_{k^\prime}}^2 \right) \right]
\no & &
+ 3 \sum_{k=1}^3
\left| \left( O_v^T O_\rho \right)_{1k} \right|^2
\left[ F \left( m_Z^2, m_{H_k}^2 \right)
  - F \left( m_W^2, m_{H_k}^2 \right) \right]
\no & & \left.
- 3 F \left( m_Z^2, m_H^2 \right) + 3 F \left( m_W^2, m_H^2 \right)
\right\},
\label{tttt}
\ea
where $m_Z$ is the $Z$-boson mass,
$m_W$ is the $W$-boson mass,
$m_H$ is the reference mass of the Higgs boson
(which is taken to be 125\,GeV),
and $s_w^2 = 1 - m_W^2 / m_Z^2$.
According to ref.~\cite{RPP},
$-0.04 < T < 0.20$.

\subsection{Extension to the quark sector}

There are many possible ways of extending our models to the quark sector.
If one envisages a model with the $CP$ symmetry~\eqref{CP},
then that symmetry must be broken spontaneously through $v_2 \neq v_3^\ast$
and that breaking must be felt in the quark sector,
because we know that there is $CP$ violation in that sector;
this can be achieved only if both scalar doublets $\phi_2$ and $\phi_3$
have Yukawa couplings to the quarks.
In a model with the interchange symmetry~\eqref{interchange},
on the other hand,
$CP$ violation may proceed through complex Yukawa couplings
and it is not necessary for $\phi_2$ and $\phi_3$ to couple to the quarks.
Things then become much simpler because
at tree level there are no flavour-changing neutral currents
mediated by the neutral scalars
and therefore the neutral scalars do not need to be so heavy.
Thus,
we extend the symmetry $\mathbbm{Z}_2^{(1)}$ of equation~\eqref{sym1} as
\be
\label{sym1new}
\mathbbm{Z}_2^{(1)}: \quad \phi_1 \to - \phi_1, \quad
D_{L \alpha} \to - D_{L \alpha}, \quad
Q_{Lk} \to - Q_{Lk},\ \forall k \in \left\{ 1, 2, 3 \right\},
\ee
where the $Q_{Lk}$ are the gauge-$SU(2)$ doublets of left-handed quarks.
With this extended $\mathbbm{Z}_2^{(1)}$,
the quarks only couple to $\phi_1$.
The Yukawa couplings of the quarks are then given by
\bs
\ba
\mathcal{L}_\mathrm{quark\, Yukawa} &=&
\sum_{\chi = u, c, t}
\overline{\chi}\, m_\chi\, \frac{- \rho_1 + i \eta_1 \gamma_5}
         {\sqrt{2} v_1}\, \chi
- \sum_{\zeta = d, s, b}
\overline{\zeta}\, m_\zeta\, \frac{\rho_1 + i \eta_1 \gamma_5}
         {\sqrt{2} v_1}\, \zeta
\no & &
+ \left[ \frac{\varphi_1^+}{v_1}\, \sum_{\chi = u, c, t}\, \sum_{\zeta = d, s, b}\,
V_{\chi \zeta}\, \overline{\chi} \left( m_\chi P_L - m_\zeta P_R \right) \zeta 
+ \mathrm{H.c.} \right]
\no &=&
- \sum_{k=1}^3 \frac{H_k \left( O_\rho \right)_{1k}}{\sqrt{2} v \sin{\beta}}
\left( \sum_{\chi = u, c, t} m_\chi\, \overline{\chi} \chi
+ \sum_{\zeta = d, s, b} m_\zeta\, \overline{\zeta} \zeta \right)
\label{vyguof} \\ & &
+ \frac{G^0 - A_3 \cot{\beta}}{\sqrt{2} v}
\left( \sum_{\chi = u, c, t} m_\chi\, \overline{\chi} i \gamma_5 \chi
- \sum_{\zeta = d, s, b} m_\zeta\, \overline{\zeta} i \gamma_5 \zeta \right)
\hspace*{8mm}
\label{gidbl} \\ & &
+ \left[ \frac{G^+ - \varphi_3^+ \cot{\beta}}{v}\,
\sum_{\chi = u, c, t}\, \sum_{\zeta = d, s, b}\,
V_{\chi \zeta}\, \overline{\chi} \left( m_\chi P_L - m_\zeta P_R \right) \zeta 
\right. \no & & \left.
+ \mathrm{H.c.} \right],
\label{bnuihop}
\ea
\es
where $P_{R,L}$ are the projectors of chirality,
$G^0$ is the neutral Goldstone boson,
$G^\pm$ are the charged Goldstone bosons,
$A_3$ is a physical pseudoscalar with mass $m_{A_3}$,
and $\varphi_3^\pm$ are the physical charged scalars with mass $m_{\varphi_3}$.
Notice in lines~\eqref{vyguof} and~\eqref{gidbl}
the absence of flavour-changing couplings of the neutral scalars.

\subsection{Procedure for producing the scatter plots}

The input for our scatter plots is $\beta$ and the eight $\lambda_p$
($p = 1, \ldots, 8$).

In order for the potential to be bounded from below
we require that the $\lambda_p$ satisfy~\cite{Kannike:2012pe}
\bs
\label{cteyik}
\ba
\lambda_1 &>& 0, \\
\lambda_2 &>& 0, \\
L_1 &>& 0, \\
L_2 &>& 0, \\
L_2 \sqrt{\lambda_1} + 2 L_1 \sqrt{\lambda_2}
- 4 \lambda_2 \sqrt{\lambda_1} + L_1 \sqrt{L_2} &>& 0.
\ea
\es
In inequalities~\eqref{cteyik},
\bs
\ba
L_1 &\equiv& 2 \sqrt{\lambda_1 \lambda_2} + \lambda_3
+ \left( \lambda_5 - 2 \left| \lambda_7 \right| \right)\,
\Theta \left( 2 \left| \lambda_7 \right| - \lambda_5 \right),
\\
L_2 &\equiv& 2 \lambda_2 + \lambda_4
+ \left( \lambda_6 - 2 \left| \lambda_8 \right| \right)
\Theta \left( 2 \left| \lambda_8 \right| - \lambda_6 \right),
\ea
\es
where $\Theta$ is the step (Heaviside) function.

In order for the potential not to break unitarity we impose
the following conditions on the $\lambda_p$,
which are derived in appendix~\ref{app:unitarity}:
\bs
\label{unita}
\ba
\left| \lambda_3 \pm \lambda_5 \right| &<& 4 \pi, \\
\left| \lambda_4 \pm \lambda_6 \right| &<& 4 \pi, \\
\left| \lambda_3 \pm 2 \lambda_7 \right| &<& 4 \pi, \\
\left| \lambda_4 \pm 2 \lambda_8 \right| &<& 4 \pi, \\
\left| \lambda_3 + 2 \lambda_5 \pm 6 \lambda_7 \right| &<& 4 \pi, \\
\left| \lambda_4 + 2 \lambda_6 \pm 6 \lambda_8 \right| &<& 4 \pi, \\
\left| 2 \lambda_2 - 2 \lambda_8 \right| &<& 4 \pi, \\
\left| 2 \lambda_2 - \lambda_6 \right| &<& 4 \pi, \\
\left| 6 \lambda_2 - 2 \lambda_4 - \lambda_6 \right| &<& 4 \pi, \\
\left| \lambda_1 + \lambda_2 + \lambda_8 \pm
\sqrt{\left( \lambda_1 - \lambda_2 - \lambda_8 \right)^2
  + 8 \lambda_7^2} \right| &<& 4 \pi, \\
\left| \lambda_1 + \lambda_2 + \frac{\lambda_6}{2} \pm
\sqrt{\left( \lambda_1 - \lambda_2 - \frac{\lambda_6}{2} \right)^2
  + 2 \lambda_5^2} \right| &<& 4 \pi, \\
\left| 3 \lambda_1 + 3 \lambda_2 + \lambda_4 + \frac{\lambda_6}{2} \right. & &
\nonumber \\ \left.
\pm \sqrt{\left( 3 \lambda_1 - 3 \lambda_2 - \lambda_4
  - \frac{\lambda_6}{2} \right)^2
  + 2 \left( 2 \lambda_3 + \lambda_5 \right)^2} \right| &<& 4 \pi.
\ea
\es

The angle $\beta$ is an input of our scatter plots.
The VEVs $v_{1,2,3}$ are determined from equations~\eqref{vbeta},
where $v$ is given by equation~\eqref{v} and $r = m_\mu / m_\tau$.
Then,
$\mu_1$,
$\mu_2$,
and $\mu_3$ are computed by using equations~\eqref{vugio}.
The value of $V_0$ is given by equation~\eqref{vvv000}.
We require $V_0 < 0$.
We also enforce a number of conditions related to the alternative
stability points in appendix~\ref{stab}:
\begin{itemize}
\item If the quantities in the right-hand sides of equations~\eqref{bvcigp}
  are both positive,
  then we require $V_0 < V_0^{(1\pm)}$,
  where the quantities $V_0^{(1\pm)}$ are given in equation~\eqref{v01}.
\item If the quantity in the right-hand side of equation~\eqref{bvnugh}
  is positive,
  then we require $V_0 < V_0^{(3)}$,
  where $V_0^{(3)}$ is given in equation~\eqref{v03}.
\item If the quantity in the right-hand side of equation~\eqref{prop}
  is positive,
  then we require $V_0 < V_0^{(4)}$,
  where $V_0^{(4)}$ is given in equation~\eqref{v04}.
\item If the quantity in the right-hand side of equation~\eqref{prop2}
  is positive (with either the plus or the minus sign),
  then we require $V_0 < V_0^{(5\pm)}$
  (with the same sign),
  where $V_0^{(5\pm)}$ are given in equation~\eqref{v05}.
\item If the quantity in the right-hand side of equation~\eqref{prop3}
  is positive and the inequality~\eqref{prop4} is satisfied,
  then we require $V_0 < V_0^{(6)}$,
  with $V_0^{(6)}$ given in equation~\eqref{v06}.
\end{itemize}

We compute the squared masses in equations~\eqref{themass}.
We construct $M_\rho$ in equation~\eqref{MR}
and diagonalize it according to equation~\eqref{diag}.

We assume that \emph{the lightest physical scalar,
\textit{viz.}~$H_1$,
corresponds to the scalar particle discovered at LHC};
we therefore fit its mass $m_{H_1}$ to be 125\,GeV.
This fit is very precise,
hence $m_{H_1}$ never needs to appear in our scatter plots.

We require that the masses of the six additional scalars,
\textit{i.e.}~$m_{\varphi_{2,3}}$,
$m_{A_{2,3}}$,
and $m_{H_{2,3}}$,
are all larger than 150\,GeV.
We also require the parameter~$T$,
computed through equation~\eqref{tttt},
to lie in between $-0.04$ and $+0.20$~\cite{RPP}.

The particle discovered at LHC,
which we interpret as our $H_1$,
couples to gauge-boson pairs,
to the heavy quarks,
and to the $\tau$ lepton with strengths close to the predictions of the SM.
We hence derive the following constraints:
\begin{itemize}
\item The strength of the coupling of $H_1$ to gauge-boson pairs,
  divided by the strength of the coupling
  of the SM Higgs boson to gauge-boson pairs,
  is~\cite{Grimus:2007if} $\left| g_{ZZ} \right|$,\footnote{The important
    quantity is $\left| g_{ZZ} \right|$,
    not $g_{ZZ}$ itself,
    because the sign of the first column of the matrix $O_\rho$ is arbitrary
    and physically meaningless,
    hence the sign of $g_{ZZ}$ is also arbitrary.
    Alternatively,
    we may reason that the physical cross sections
    depend on the squared amplitudes,
    hence on $g_{ZZ}^2$,
    not on the amplitudes themselves.}
  where
  \be \label{gZZ}
  g_{ZZ} \equiv \frac{1}{v}\, \sum_{k=1}^3 v_k \left( O_\rho \right)_{k1}.
  \ee
  Note that $-1 \le g_{ZZ} \le 1$,
  because $g_{ZZ}$ is the scalar product of two unit vectors.
  The limit $\left| g_{ZZ} \right| = 1$
  corresponds to $H_1$ coupling to pairs of gauge bosons
  with exactly the same strength as the SM Higgs boson does.
  In our scatter plots we require
  \be
  \left| g_{ZZ} \right| > 0.9.
  \ee
\item We observe in equation~\eqref{vyguof} that
  $H_1$ couples to the quarks with strength
  $\left. \left( O_\rho \right)_{11} \right/ \! \sin{\beta}$
  times the strength of the coupling to the quarks of the SM Higgs boson.
  Since the sign of $\left( O_\rho \right)_{11}$ is physically meaningless
  but is correlated with the sign of $g_{ZZ}$,
  we define
  \be
  \label{gquarks}
  g_\mathrm{quarks} \equiv \frac{\left( O_\rho \right)_{11}}{\sin{\beta}}\,
  \frac{g_{ZZ}}{\left| g_{ZZ} \right|}.
  \ee
  In our scatter plots we demand that $0.9 < g_\mathrm{quarks} < 1.1$.
\item We use $\left| v_2 / v_3 \right| = m_\mu / m_\tau$;
  this means that we are assuming that,
  in our specific model,
  it is the scalar doublet $\phi_3$ that couples to $\overline{D_{L\tau}} \tau_R$.
  Thus,
  there is a Yukawa coupling
  \be
  \Upsilon\, \overline{\tau_L} \tau_R
  \left( v_3 + \frac{\rho_3 + i \eta_3}{\sqrt{2}} \right) + \mathrm{H.c.}
  \ee
  The modulus of the Yukawa coupling constant $\Upsilon$
  of course is $m_\tau / v_3$.
  Since
  \be
  \rho_3 = \sum_{k=1}^3 \left( O_\rho \right)_{3k} H_k,
  \ee
  $H_1$ couples to $\overline{\tau_L} \tau_R$
  with strength $\left( O_\rho \right)_{31}
  \left( \Upsilon \left/ \sqrt{2} \right. \right)$.
  The modulus of the coupling of the SM Higgs boson
  to $\overline{\tau_L} \tau_R$
  is $m_\tau \left/ \left( \sqrt{2} v \right) \right.$.
  Therefore,
  for $H_1$ to couple to $\tau$ leptons with the same strength
  as the SM Higgs boson,
  one needs to have
  $\left| \left. \left( O_\rho \right)_{31} \right/ \! v_3 \right| \approx 1/v$.
  Defining
  \be
  \label{gtau}
  g_\tau \equiv \frac{\left( O_\rho \right)_{31} v}{v_3}\,
  \frac{g_{ZZ}}{\left| g_{ZZ} \right|},
  \ee
  we demand that $0.9 < g_\tau < 1.1$.
\end{itemize}

Furthermore,
we see in equation~\eqref{bnuihop} that the physical charged scalars
$\varphi_3^\pm$ interact with the quarks in the same way
as the charged scalars of the type-I two-Higgs-doublet model.
Therefore,
in our scatter plots
we have borrowed the bounds in the $\tan{\beta}$--$m_{\varphi_3}$ plane
given in figure~18 of ref.~\cite{2HDM}.

\subsection{Scatter plots}

\begin{figure}
\begin{center}
\epsfig{file=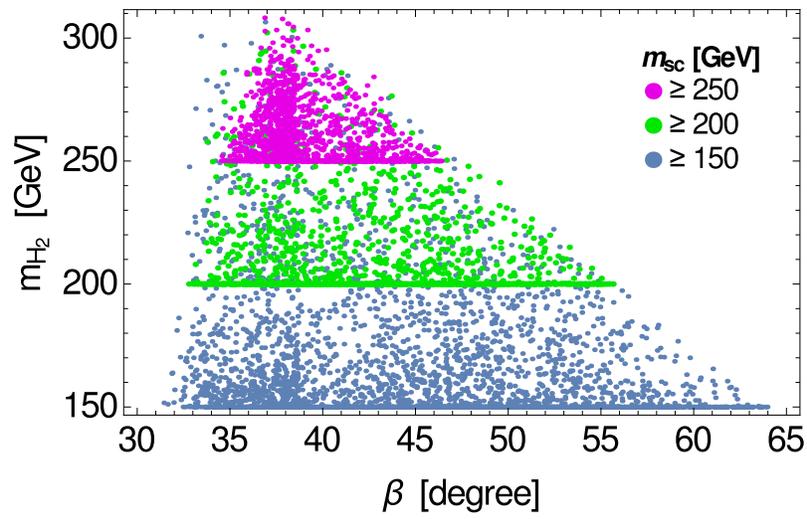,width=0.8\textwidth}
\end{center}
\caption{Scatter plot of $m_{H_2}$ \textit{versus} the angle $\beta$.
  Blue points have all the scalar masses,
  except $m_{H_1} = 125\, \mathrm{GeV}$,
  higher than 150\,GeV;
  green points have all those masses higher than 200\,GeV,
  and magenta points have all of them higher than 250\,GeV.
  \label{fig1}}
\end{figure}

In figure~\ref{fig1}
we plot the mass of the lightest new scalar,
\textit{i.e.}\ of $H_2$,
against $\beta$.
One sees that $\beta$ must always be close to $45^\circ$
and that $\beta$ becomes ever more restricted
when the new-scalar masses get higher.
Also notice that $m_{H_2}$ cannot be much higher than 300\,GeV.

\begin{figure}
\begin{center}
\epsfig{file=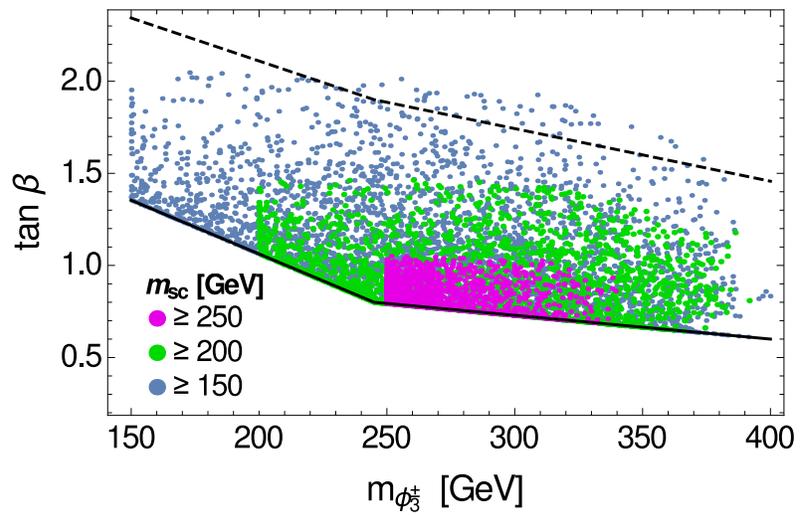,width=0.8\textwidth}
\end{center}
\caption{Scatter plot of $\tan{\beta}$ \textit{versus}
  the mass of the physical charged scalars $\varphi_3^\pm$.
  The notation for the colours is the same as in figure~\ref{fig1}.
  The solid line and the dashed line are
  phenomenological bounds extracted from figure~18 of ref.~\cite{2HDM}.
  \label{fig2}}
\end{figure}

In figure~\ref{fig2}
we plot $\tan{\beta}$ against the mass
of the physical charged scalars $\varphi_3^\pm$
that interact with the quarks.
Also marked in figure~\ref{fig2},
through a solid line,
is the phenomenological lower bound on the mass of $\varphi_3^\pm$,
which we have taken from figure~18 of ref.~\cite{2HDM}.
That bound incorporates the constraints from $Z \to b \bar b$,
$\epsilon_K$,
and $\Delta m_{B_s}$;
it guarantees that the charged scalars $\varphi_3^\pm$
do not mediate excessively strong $\left| \Delta S \right| = 2$ transitions
through box diagrams.

\begin{figure}
\begin{center}
\epsfig{file=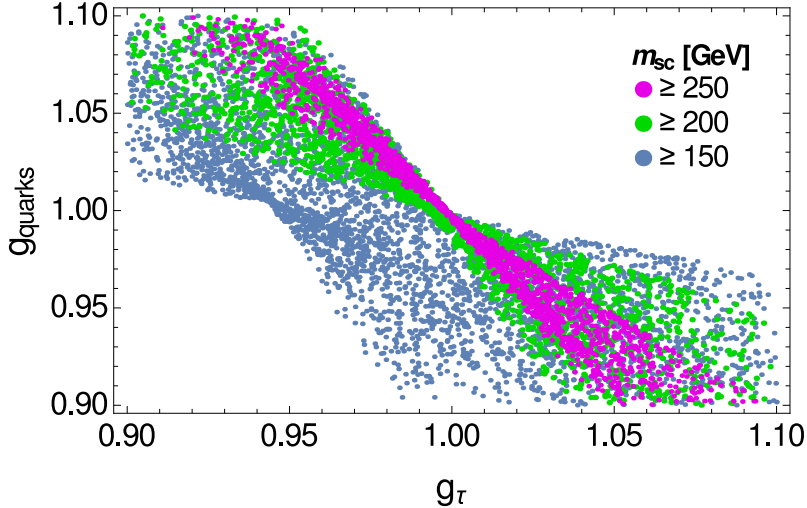,width=0.8\textwidth}
\end{center}
\caption{Scatter plot of $g_\mathrm{quarks}$ \textit{versus} $g_\tau$.
  The notation for the colours of the points
  is the same as in figure~\ref{fig1}.
  \label{fig3}}
\end{figure}

In figure~\ref{fig3}
we plot the quantities defined in equations~\eqref{gquarks} and~\eqref{gtau}
against each other.
They seem to be anti-correlated;
the anti-correlation becomes more well-defined
when the masses of all the new scalar particles are higher.

\begin{figure}
\begin{center}
\epsfig{file=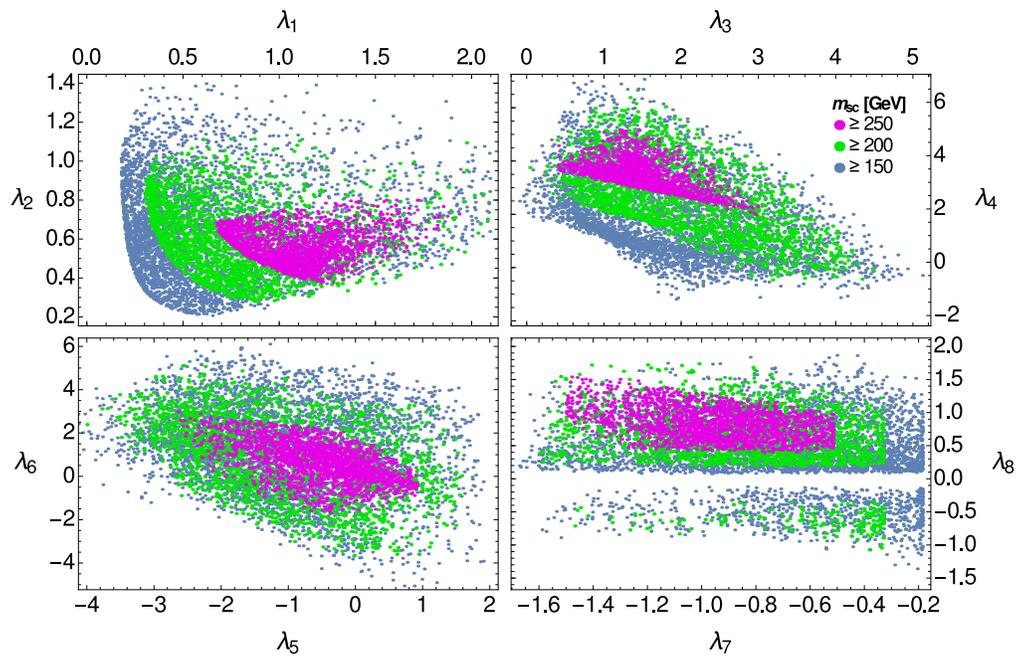,width=1.0\textwidth}
\end{center}
\caption{Scatter plots of the $\lambda_p$.
  The notation for the colours of the points
  is the same as in figure~\ref{fig1}.
  \label{fig4}}
\end{figure}

In figure~\ref{fig4}
we plot the eight parameters $\lambda_p$ of the scalar potential.
One observes that $\left| \lambda_p \right|$ is never larger than 2
for $p \in \left\{ 1, 2, 7, 8 \right\}$;
for $3 \le p \le 6$ the $\lambda_p$ may be somewhat larger.

\begin{figure}
\begin{center}
\epsfig{file=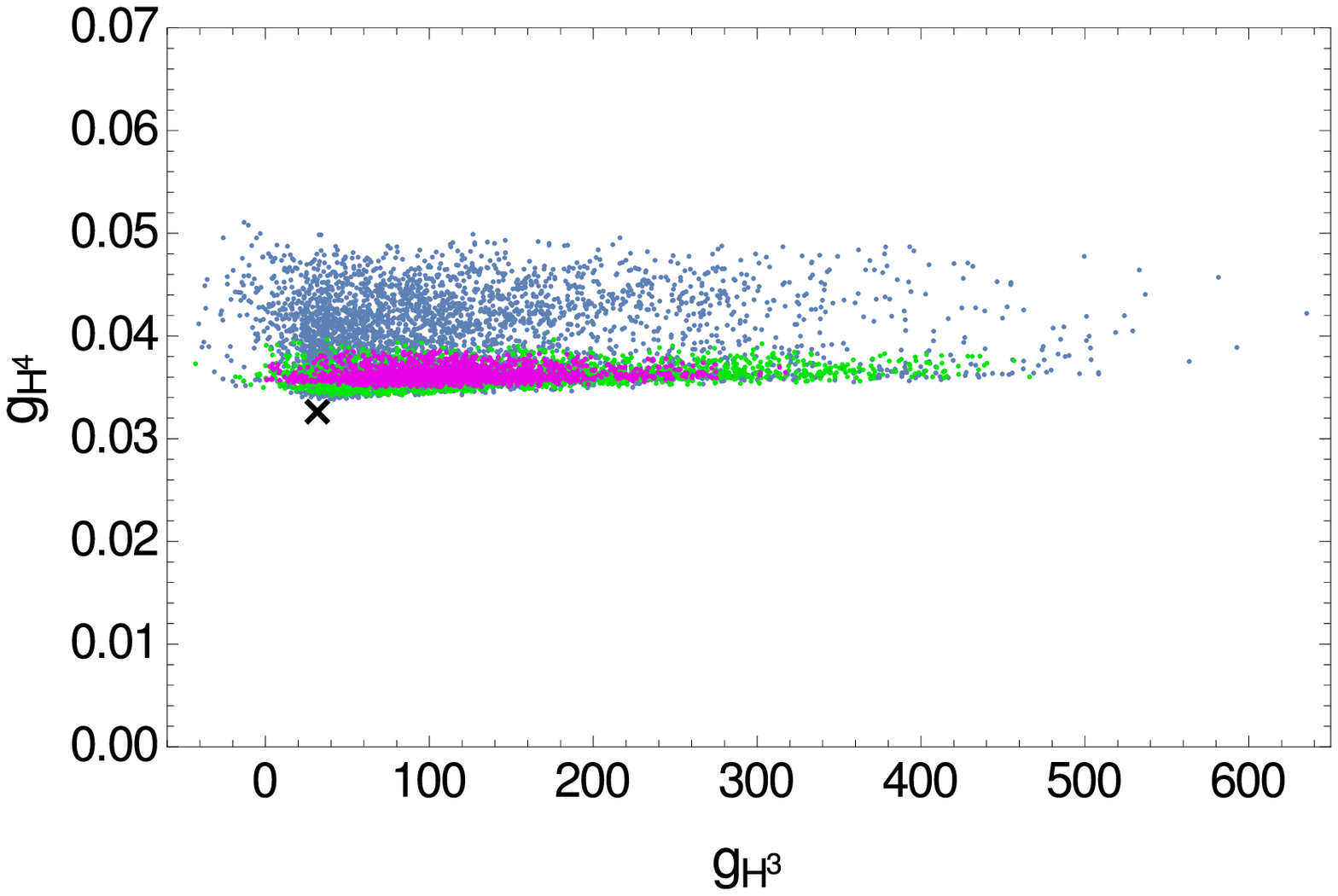,width=0.8\textwidth}
\end{center}
\caption{Scatter plot of $g_{H^4}$ \textit{versus} $g_{H^3}$.
  The notation for the colours of the points
  is the same as in figure~\ref{fig1}.
  The black cross indicates the values of $g_{H^4}$ and $g_{H^3}$ in the SM.
  \label{fig5}}
\end{figure}

In figure~\ref{fig5} we have plotted
the quartic Higgs coupling $g_{H^4}$ against the cubic Higgs coupling $g_{H^3}$.
These are the coefficients of the terms
$\left( H_1 \right)^4$ and $\left( H_1 \right)^3$,
respectively,
in the Lagrangian;
in the case of $g_{H^3}$ we have multiplied the coefficient
of $\left( H_1 \right)^3$ by $g_{ZZ} \left/ \left| g_{ZZ} \right| \right.$
in order to take into account the possibility that the field $H_1$
has the wrong sign.
One sees that the three-Higgs coupling may be almost twenty times larger
than in the SM.
Also,
that coupling may be zero or even negative,
\textit{i.e.}\ it may have a sign opposite to the one in the SM.
The four-Higgs coupling is always larger than the corresponding SM coupling;
it may at most be 60\% larger than in the SM.
We point out that,
in a general two-Higgs-doublet-model,
the three-Higgs coupling has less freedom
(it may at most be ten times larger than in the SM) than in this model,
while the four-Higgs coupling has much more freedom
than in this model---it may have values from zero
until almost fifteen times larger than in the SM~\cite{newPaper}.
Therefore,
a measurement of $g_{H^3}$---of the cubic interaction of the 125\,GeV
scalar---may produce a large surprise
and even distinguish this three-Higgs-doublet model
from the most general two-Higgs-doublet one.

\begin{figure}
\begin{center}
\epsfig{file=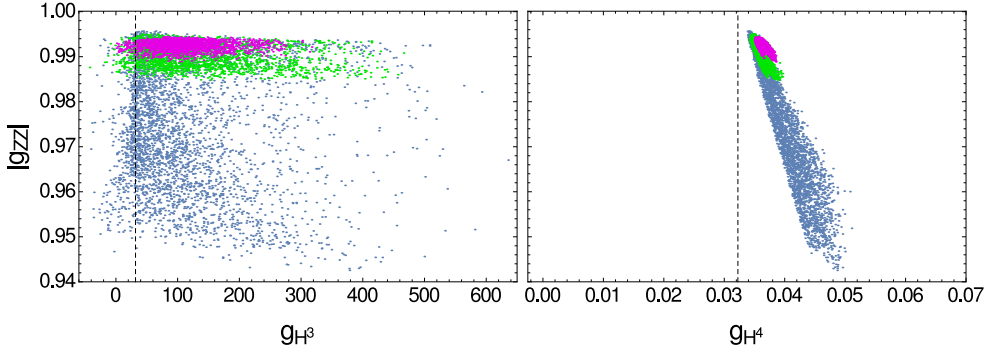,width=1.0\textwidth}
\end{center}
\caption{Scatter plots of $\left| g_{ZZ} \right|$ \textit{versus}
  $g_{H^3}$ and $g_{H^4}$.
  The notation for the colours of the points
  is the same as in figure~\ref{fig1}.
  The dashed vertical lines indicate the values of
  $g_{H^3}$ and $g_{H^4}$ in the SM.
  \label{fig6}}
\end{figure}

In figure~\ref{fig6}
we have plotted $\left| g_{ZZ} \right|$ against the quartic Higgs coupling
and against the cubic Higgs coupling.
Notice that,
although in our search we have restricted $\left| g_{ZZ} \right|$
to have values in the range from 0.9 to 1,
we have ended up obtaining only points
with $\left| g_{ZZ} \right| > 0.94$.
This is because we have restricted all the scalar masses
(except the one of $H_1$)
to be larger than 150\,GeV;
larger scalar masses require a larger $\left| g_{ZZ} \right|$
because the values of $\left| g_{ZZ} \right|$ approach unity
when the masses of all the new scalars
are higher---this is the decoupling limit.

\section{Conclusions} \label{sec:conclusions}

In this paper we have constructed various extensions of the SM
that yield predictions for the effective light-neutrino Majorana mass matrix $M$
given in terms of the charged-lepton mass ratios.
We have produced twelve models $M_{\alpha p q}$,
where $\alpha \in \left\{ e,\ \mu,\ \tau \right\}$
and $p, q \in \left\{ 1,\ 2 \right\}$.
Models $M_{\alpha 1 q}$ predict
\be
\label{bnih}
\left| \frac{M_{\beta \beta}}{M_{\gamma \gamma}} \right| = \frac{m_\beta^2}{m_\gamma^2},
\quad
\left| \frac{M_{\alpha \beta}}{M_{\alpha \gamma}} \right| = \frac{m_\beta}{m_\gamma},
\ee
where $\alpha \neq \beta \neq \gamma \neq \alpha$,
whereas models $M_{\alpha 2 q}$ predict
\be
\label{bnih2}
\left| \frac{M_{\beta \beta}}{M_{\gamma \gamma}} \right| = \frac{m_\gamma^2}{m_\beta^2},
\quad
\left| \frac{M_{\alpha \beta}}{M_{\alpha \gamma}} \right| = \frac{m_\gamma}{m_\beta}.
\ee
Furthermore,
models $M_{\alpha p 1}$ predict
\be
\label{hui}
\arg{\left[ M_{\gamma \gamma} \left( M_{\alpha \beta} \right)^2
    M^\ast_{\beta \beta} \left( M^\ast_{\alpha \gamma} \right)^2 \right]} = 0,
\ee
whereas models $M_{\alpha p 2}$ predict
\bs
\label{buih}
\ba
\arg{\left[ M^\ast_{\beta \beta} M^\ast_{\gamma \gamma} \left( M_{\beta \gamma} \right)^2
    \right]} &=& 0,
\\
\arg{\left( M^\ast_{\alpha \alpha} M^\ast_{\beta \gamma}
    M_{\alpha \beta} M_{\alpha \gamma}
    \right)} &=& 0.
\ea
\es
In practice,
the conditions~\eqref{hui} or~\eqref{buih} are not so important;
this is because conditions~\eqref{bnih} or~\eqref{bnih2}
mean that two matrix elements of $M$ are relatively small
and lead to our models being approximations to two-texture-zero cases.
Thus,
eight of our twelve models are able to correctly fit the data:
\begin{itemize}
\item Models $M_{\mu 1 q}$ for $q = 1, 2$,
  which are approximations to case~A$_1$.
\item Models $M_{\tau 1 q}$ for $q = 1, 2$,
  which are approximations to case~A$_2$.
\item Models $M_{e 1 q}$ for $q = 1, 2$,
  which are approximations to case~B$_3$.
\item Models $M_{e 2 q}$ for $q = 1, 2$,
  which are approximations to case~B$_4$.
\end{itemize}
The four models~$M_{\mu 2 q}$ and $M_{\tau 2 q}$
are not compatible with the phenomenological data
and are therefore excluded.

We have emphasized that our models $M_{e1q}$ lead,
just from the two conditions
\be
\left| \frac{M_{\mu\mu}}{M_{\tau\tau}} \right| = \frac{m_\mu^2}{m_\tau^2},
\quad
\left| \frac{M_{e\mu}}{M_{e\tau}} \right| = \frac{m_\mu}{m_\tau},
\ee
to a vast predictive power,
\textit{viz.}\ $\delta \approx 3 \pi / 2$,
$\alpha_{21} \approx 0$,
$\alpha_{31} \approx \pi$,
and almost maximal neutrinoless double-beta decay
for either a normal or an inverted neutrino mass spectrum.
Moreover,
the quadrant of $\theta_{23}$ is correlated with the type of mass spectrum
and $\theta_{23}$ approaches $\pi/4$ when the neutrino masses increase.

We have carefully worked out a scalar potential
appropriate to our models $M_{e1q}$.
(With slight modifications and no qualitatively different results,
the potential is also appropriate to models $M_{\mu 1q}$ and $M_{\tau 1q}$.)
Our assumptions were the following:
\begin{itemize}
\item There are only three Higgs doublets $\phi_{1,2,3}$.
\item There is an interchange symmetry $\phi_2 \leftrightarrow \phi_3$
  that is \emph{not}\/ softly broken in the quadratic part
  of the scalar potential.
\item The potential has
  an unbroken symmetry under $\phi_1 \to - \phi_1$.
\item The vacuum expectation values are real.
\item The symmetry $\phi_1 \to - \phi_1$ is extended to the quark sector
  in such a way that only $\phi_1$ has Yukawa couplings to the quarks;
  the physical neutral scalars therefore have
  no flavour-changing Yukawa couplings.
  $CP$ violation is \emph{hard},
  \textit{i.e.}\ it originates in complex Yukawa couplings.
\item The particle with mass 125\,GeV discovered at LHC
  is \emph{the lightest}\/ physical scalar.
\end{itemize}
Through a careful simulation we have found the appropriate ranges
for the various parameters of the scalar potential.
The physical-scalar masses cannot be much higher than a few hundred GeV.

\vspace*{5mm}

\paragraph{Acknowledgements:}
L.L.~thanks Pedro M.~Ferreira,
Jo\~ao Paulo Silva,
and Igor Ivanov for useful discussions.
D.J.~thanks the Lithuanian Academy of Sciences
for support through the project~DaFi2017.
The work of L.L.\ is supported by the Portuguese
\textit{Fun\-da\-\c c\~ao pa\-ra a Ci\-\^en\-cia e a Te\-cno\-lo\-gia}\/
through the projects CERN/FIS-NUC/0010/2015
and UID/FIS/00777/2013,
which are partially funded by POCTI (FEDER),
COMPETE,
QREN,
and the European Union.

\newpage

\begin{appendix}

\setcounter{equation}{0}
\renewcommand{\theequation}{A\arabic{equation}}

\section{Unitarity bounds for a 3HDM with
  $\mathbbm{Z}_2 \times \mathbbm{Z}_2 \times \mathbbm{Z}_2$ symmetry}
\label{app:unitarity}

\subsection{General case}

We consider the most general three-Higgs-doublet model
with $\mathbbm{Z}_2^{(1)} \times \mathbbm{Z}_2^{(2)}
\times \mathbbm{Z}_2^{(3)}$ symmetry,
where
\be
\label{hfyti}
\mathbbm{Z}_2^{(1)}:\ \phi_1 \to - \phi_1; \quad
\mathbbm{Z}_2^{(2)}:\ \phi_2 \to - \phi_2; \quad
\mathbbm{Z}_2^{(3)}:\ \phi_3 \to - \phi_3.
\ee
It is immaterial in this appendix whether
any of the symmetries~\eqref{hfyti} is softly broken;
here we just deal with the quartic part of the potential
\bs
\label{V}
\ba
V_\mathrm{quartic} &=&
\Lambda_1 \left( \phi_1^\dagger \phi_1 \right)^2
+ \Lambda_2 \left( \phi_2^\dagger \phi_2 \right)^2
+ \Lambda_3 \left( \phi_3^\dagger \phi_3 \right)^2
\\ & &
+ \Lambda_4\, \phi_1^\dagger \phi_1\, \phi_2^\dagger \phi_2
+ \Lambda_5\, \phi_1^\dagger \phi_1\, \phi_3^\dagger \phi_3
+ \Lambda_6\, \phi_2^\dagger \phi_2\, \phi_3^\dagger \phi_3
\\ & &
+ \Lambda_7\, \phi_1^\dagger \phi_2\, \phi_2^\dagger \phi_1
+ \Lambda_8\, \phi_1^\dagger \phi_3\, \phi_3^\dagger \phi_1
+ \Lambda_9\, \phi_2^\dagger \phi_3\, \phi_3^\dagger \phi_2
\\ & &
+ \left[ \Lambda_{10} \left( \phi_1^\dagger \phi_2 \right)^2
+ \Lambda_{11} \left( \phi_1^\dagger \phi_3 \right)^2
+ \Lambda_{12} \left( \phi_2^\dagger \phi_3 \right)^2
+ \mathrm{H.c.} \right], \hspace*{6mm}
\ea
\es
where $\Lambda_{1,\ldots,9}$ are real
and $\Lambda_{10,11,12}$ are in general complex.
We follow ref.~\cite{Bento:2017eti} to compute the unitarity bounds
on the parameters of the potential~\eqref{V}.
For notational simplicity,
we write
\be
\label{abc}
\phi_1 = \left( \begin{array}{c} a \\ b \end{array} \right), \quad
\phi_2 = \left( \begin{array}{c} c \\ d \end{array} \right), \quad
\phi_3 = \left( \begin{array}{c} e \\ f \end{array} \right),
\ee
where the letters $a, \ldots, f$ denote creation/destruction operators
as well as the corresponding particles.
The (non-)existence of vacuum expectation values
is immaterial for the unitarity bounds,
therefore we neglect them in the notation~\eqref{abc}.
We denote the Hermitian-conjugate operators through bars:
$a^\dagger \to \ab$,
$b^\dagger \to \bb$,
and so on.
Then,
\bs
\ba
V_\mathrm{quartic} &=&
\Lambda_1 \left( \ab \ab a a + \bb \bb b b + 2 \ab \bb a b \right)
\\ & &
+ \Lambda_2 \left( \cb \cb c c + \db \db d d + 2 \cb \db c d \right)
\\ & &
+ \Lambda_3 \left( \eb \eb e e + \fb \fb f f + 2 \eb \fb e f \right)
\\ & &
+ \Lambda_4 \left( \ab \cb a c + \bb \db b d + \ab \db a d + \bb \cb b c \right)
\\ & &
+ \Lambda_5 \left( \ab \eb a e + \bb \fb b f + \ab \fb a f + \bb \eb b e \right)
\\ & &
+ \Lambda_6 \left( \cb \eb c e + \db \fb d f + \cb \fb c f + \db \eb d e \right)
\\ & &
+ \Lambda_7 \left( \ab \cb a c + \bb \db b d + \ab \db b c + \bb \cb a d \right)
\\ & &
+ \Lambda_8 \left( \ab \eb a e + \bb \fb b f + \ab \fb b e + \bb \eb a f \right)
\\ & &
+ \Lambda_9 \left( \cb \eb c e + \db \fb d f + \cb \fb d e + \db \eb c f \right)
\\ & &
+ \Lambda_{10} \left( \ab \ab c c + \bb \bb d d + 2 \ab \bb c d \right)
\\ & &
+ \Lambda_{10}^\ast \left( \cb \cb a a + \db \db b b + 2 \cb \db a b \right)
\\ & &
+ \Lambda_{11} \left( \ab \ab e e + \bb \bb f f + 2 \ab \bb e f \right)
\\ & &
+ \Lambda_{11}^\ast \left( \eb \eb a a + \fb \fb b b + 2 \eb \fb a b \right)
\\ & &
+ \Lambda_{12} \left( \cb \cb e e + \db \db f f + 2 \cb \db e f \right)
\\ & &
+ \Lambda_{12}^\ast \left( \eb \eb c c + \fb \fb d d + 2 \eb \fb c d \right).
\ea
\es
We must consider all the $2 \to 2$ scatterings
that various pairs of particles may suffer among themselves.
For instance,
the three states $aa$,
$cc$,
and $ee$ may,
at tree-level,
scatter through a matrix
  \be
  \label{gjguyti}
\left( \begin{array}{ccc}
2 \Lambda_1 & 2 \Lambda_{10} & 2 \Lambda_{11} \\
2 \Lambda_{10}^\ast & 2 \Lambda_2 & 2 \Lambda_{12} \\
2 \Lambda_{11}^\ast & 2 \Lambda_{12}^\ast & 2 \Lambda_3
\end{array} \right).
\ee
The scattering matrices of the states $\left( ad,\, bc \right)$,
$\left( af,\, be \right)$,
and $\left( bc,\, de \right)$ are
\be
\left( \begin{array}{cc}
  \Lambda_4 & \Lambda_7 \\ \Lambda_7 & \Lambda_4
\end{array} \right), \quad
\left( \begin{array}{cc}
  \Lambda_5 & \Lambda_8 \\ \Lambda_8 & \Lambda_5
\end{array} \right), \quad
\left( \begin{array}{cc}
  \Lambda_6 & \Lambda_9 \\ \Lambda_9 & \Lambda_6
\end{array} \right),
\ee
respectively.
The scattering matrices of the states $\left( a\db,\, \bb c \right)$,
$\left( a\fb,\, \bb e \right)$,
and $\left( \fb c,\, \db e \right)$ are
\be
\label{bihop}
\left( \begin{array}{cc}
  \Lambda_4 & 2 \Lambda_{10} \\ 2 \Lambda_{10}^\ast & \Lambda_4
\end{array} \right), \quad
\left( \begin{array}{cc}
  \Lambda_5 & 2 \Lambda_{11} \\ 2 \Lambda_{11}^\ast & \Lambda_5
\end{array} \right), \quad
\left( \begin{array}{cc}
  \Lambda_6 & 2 \Lambda_{12} \\ 2 \Lambda_{12}^\ast & \Lambda_6
\end{array} \right),
\ee
respectively.
The scattering matrix of the states $\left( a\bb,\, c\db,\, e\fb \right)$ is
\be
\label{bughoip}
\left( \begin{array}{ccc}
  2 \Lambda_1 & \Lambda_7 & \Lambda_8 \\
  \Lambda_7 & 2 \Lambda_2 & \Lambda_9 \\
  \Lambda_8 & \Lambda_9 & 2 \Lambda_3
\end{array} \right).
\ee
The scattering matrix of the states
$\left( \ab a,\, \bb b,\, \cb c,\, \db d,\, \eb e,\, \fb f \right)$ is
\be
\label{bvuifpl}
\left( \begin{array}{cccccc}
  4 \Lambda_1 & 2 \Lambda_1 & \Lambda_4 + \Lambda_7 & \Lambda_4 &
  \Lambda_5 + \Lambda_8 & \Lambda_5 \\
  2 \Lambda_1 & 4 \Lambda_1 & \Lambda_4 & \Lambda_4 + \Lambda_7 &
  \Lambda_5 & \Lambda_5 + \Lambda_8 \\
  \Lambda_4 + \Lambda_7 & \Lambda_4 & 4 \Lambda_2 & 2 \Lambda_2 & 
  \Lambda_6 + \Lambda_9 & \Lambda_6 \\
  \Lambda_4 & \Lambda_4 + \Lambda_7 & 2 \Lambda_2 & 4 \Lambda_2 & 
  \Lambda_6 & \Lambda_6 + \Lambda_9 \\
  \Lambda_5 + \Lambda_8 & \Lambda_5 & 
  \Lambda_6 + \Lambda_9 & \Lambda_6 & 4 \Lambda_3 & 2 \Lambda_3 \\
  \Lambda_5 & \Lambda_5 + \Lambda_8 & 
  \Lambda_6 & \Lambda_6 + \Lambda_9 & 2 \Lambda_3 & 4 \Lambda_3
\end{array} \right).
\ee
In order to guarantee unitarity,
we must enforce the condition that
\emph{the moduli of}\/ all the eigenvalues of these matrices
(and of a few more analogous matrices)
are smaller than $4 \pi$.
After some effort we find that those eigenvalues are
\bs
\label{fygfirp}
\ba
\Lambda_4 \pm \Lambda_7, &  
\Lambda_5 \pm \Lambda_8, &
\Lambda_6 \pm \Lambda_9, \\
\Lambda_4 \pm 2 \left| \Lambda_{10} \right|, &
\Lambda_5 \pm 2 \left| \Lambda_{11} \right|, &
\Lambda_6 \pm 2 \left| \Lambda_{12} \right|, \\
\Lambda_4 + 2 \Lambda_7 \pm 6 \left| \Lambda_{10} \right|, &
\Lambda_5 + 2 \Lambda_8 \pm 6 \left| \Lambda_{11} \right|, &
\Lambda_6 + 2 \Lambda_9 \pm 6 \left| \Lambda_{12} \right|, \hspace*{6mm}
\ea
\es
and the eigenvalues of the matrices~\eqref{gjguyti},
\eqref{bughoip},
and
\be
\label{vpgoit}
\left( \begin{array}{ccc}
  6 \Lambda_1 & 2 \Lambda_4 + \Lambda_7 & 2 \Lambda_5 + \Lambda_8 \\
  2 \Lambda_4 + \Lambda_7 & 6 \Lambda_2 & 2 \Lambda_6 + \Lambda_9 \\
  2 \Lambda_5 + \Lambda_8 & 2 \Lambda_6 + \Lambda_9 & 6 \Lambda_3
\end{array} \right).
\ee

\subsection{Case with additional symmetry $\phi_2 \leftrightarrow \phi_3$}

In our case there is an additional symmetry $\phi_2 \leftrightarrow \phi_3$
in the potential,
and that simplifies things much.
Comparing equations~\eqref{vv} and~\eqref{V},
we see that
\bs
\ba
\Lambda_1 &\to& \lambda_1, \\
\Lambda_2, \Lambda_3 &\to& \lambda_2, \\
\Lambda_4, \Lambda_5 &\to& \lambda_3, \\
\Lambda_6 &\to& \lambda_4, \\
\Lambda_7, \Lambda_8 &\to& \lambda_5, \\
\Lambda_9 &\to& \lambda_6, \\
\Lambda_{10}, \Lambda_{11} &\to& \lambda_7, \\
\Lambda_{12} &\to& \lambda_8.
\ea
\es
The quantities~\eqref{fygfirp} then become
\bs
\ba
\lambda_3 \pm \lambda_5, & & \lambda_4 \pm \lambda_6, \\
\lambda_3 \pm 2 \lambda_7, & & \lambda_4 \pm 2 \lambda_8, \\
\lambda_3 + 2 \lambda_5 \pm 6 \lambda_7, & &
\lambda_4 + 2 \lambda_6 \pm 6 \lambda_8,
\ea
\es
and the matrices~\eqref{gjguyti},
\eqref{bughoip},
and~\eqref{vpgoit} become
\be
\begin{array}{c}
  \left( \begin{array}{ccc}
2 \lambda_1 & 2 \lambda_7 & 2 \lambda_7 \\
2 \Lambda_7 & 2 \lambda_2 & 2 \lambda_8 \\
2 \lambda_7 & 2 \lambda_8 & 2 \lambda_2
\end{array} \right),
\quad
\left( \begin{array}{ccc}
  2 \lambda_1 & \lambda_5 & \lambda_5 \\
  \lambda_5 & 2 \lambda_2 & \lambda_6 \\
  \lambda_5 & \lambda_6 & 2 \lambda_2
\end{array} \right),
\\*[7mm]
\left( \begin{array}{ccc}
  6 \lambda_1 & 2 \lambda_3 + \lambda_5 & 2 \lambda_3 + \lambda_5 \\
  2 \lambda_3 + \lambda_8 & 6 \lambda_2 & 2 \lambda_4 + \lambda_6 \\
  2 \lambda_3 + \lambda_5 & 2 \lambda_4 + \lambda_6 & 6 \lambda_2
\end{array} \right).
\end{array}
\label{mats}
\ee
The matrices~\eqref{mats} are 2--3 symmetric and therefore their
eigenvalues are easy to find.
One thus obtains the quantities in the left-hand sides
of inequalities~\eqref{unita}.

\setcounter{equation}{0}
\renewcommand{\theequation}{B\arabic{equation}}

\section{Other stability points} \label{stab}

Besides the vacuum state given by equations~\eqref{vugio} and~\eqref{vvv000},
there are several other stability points of the potential.
The vacuum state must have a lower value of the potential
than all other stability points.
Therefore we must consider as many stability points as we can and,
for each of them,
compute the expectation value of the potential.
That is what we do in the following.
\begin{enumerate}
\item Equations~\eqref{vs} have solutions with $v_1 \neq 0$ and $v_3 = \pm v_2$.
  They are
  \bs
  \label{bvcigp}
  \ba
  v_1^2 &=& \frac{2 \left( \lambda_2 + l_4 \right) \mu_1
    - 4 l_3 \left( \mu_2 \pm \mu_3 \right)}{8 l_3^2
    - 4 \lambda_1 \left( \lambda_2 + l_4 \right)},
  \\
  v_2^2 &=& \frac{- 2 l_3 \mu_1
    + 2 \lambda_1 \left( \mu_2 \pm \mu_3 \right)}{8 l_3^2
    - 4 \lambda_1 \left( \lambda_2 + l_4 \right)}.
  \ea
  \es
  Plugging $v_3 = \pm v_2$
  together with equations~\eqref{bvcigp} into equation~\eqref{v0},
  one obtains
  \be
  \label{v01}
  V_0 = V_0^{(1\pm)} \equiv \frac{\left( \lambda_2 + l_4 \right) \mu_1^2
    + 2 \lambda_1 \left( \mu_2 \pm \mu_3 \right)^2
    - 4 l_3 \mu_1 \left( \mu_2 \pm \mu_3 \right)}{8 l_3^2
    - 4 \lambda_1 \left( \lambda_2 + l_4 \right)}.
  \ee
\item The point $v_1 = v_2 = v_3 = 0$ has
  \be
  \label{v02}
  V_0 = V_0^{(2)} \equiv 0.
  \ee
\item If $v_2 = v_3 = 0$ but $v_1 \neq 0$,
  there is a stability point with
  \be
  \label{bvnugh}
  v_1^2 = - \frac{\mu_1}{2 \lambda_1},
  \ee
  yielding
  \be
  \label{v03}
  V_0 = V_0^{(3)} \equiv - \frac{\mu_1^2}{4 \lambda_1}.
  \ee
\item If $v_1 = 0$ but $v_2\neq 0$ and $v_3 \neq 0$,
  we may analytically entertain the possibility that
  the VEVs of $\phi_2^0$ and $\phi_3^0$ have a relative phase $\vartheta$.
  We take in this case both $v_2$ and $v_3$ to be \emph{positive}\/ and
  \ba
  V_0 &=& \mu_2 \left( v_2^2 + v_3^2 \right) +
  \lambda_2 \left( v_2^4 + v_3^4 \right) +
  \left( \lambda_4 + \lambda_6 \right) v_2^2 v_3^2
  \no & &
  + 2 \mu_3 v_2 v_3 \cos{\vartheta}
  + 2 \lambda_8 v_2^2 v_3^2 \cos{\left( 2 \vartheta \right)}.
  \label{vo3}
  \ea
  The stationarity equations are
  \ba
  0 &=& \mu_3 \sin{\vartheta} + 2 \lambda_8 v_2 v_3
  \sin{\left( 2 \vartheta \right)},
  \no
  0 &=& \mu_2 v_2 + \mu_3 v_3 \cos{\vartheta}
  + 2 \lambda_2 v_2^3 + \left( \lambda_4 + \lambda_6 \right) v_2 v_3^2
  + 2 \lambda_8 v_2 v_3^2 \cos{\left( 2 \vartheta \right)},
  \no
  0 &=& \mu_2 v_3 + \mu_3 v_2 \cos{\vartheta}
  + 2 \lambda_2 v_3^3 + \left( \lambda_4 + \lambda_6 \right) v_2^2 v_3
  + 2 \lambda_8 v_2^2 v_3 \cos{\left( 2 \vartheta \right)}.
  \no
  \label{h3}
  \ea
  This leads to the following possibilities:
  \begin{enumerate}
  \item $\cos{\vartheta} = \pm 1$ and $v_3 \neq v_2$.
    Then,
    \bs
    \label{bkfpgo}
    \ba
    v_2^2 + v_3^2 &=& - \frac{\mu_2}{2 \lambda_2}, \label{prop}
    \\
    \left( 2 \lambda_2 - \lambda_4 - \lambda_6 - 2 \lambda_8
    \right) v_2 v_3 &=& \pm \mu_3.
    \ea
    \es
    Plugging $\cos{\vartheta} = \pm 1$ and equations~\eqref{bkfpgo}
    into equation~\eqref{vo3},
    one obtains
    \be
    \label{v04}
    V_0 = V_0^{(4)} \equiv
    - \frac{\mu_2^2}{4 \lambda_2} + \frac{\mu_3^2}{2 \lambda_2
      - \lambda_4 - \lambda_6 - 2 \lambda_8}.
    \ee
  \item $\cos{\vartheta} = \pm 1$ and $v_3 = v_2$.
    One then has
    \be
    \label{prop2}
    v_2^2 = - \frac{\mu_2 \pm \mu_3}{2 \lambda_2
      + \lambda_4 + \lambda_6 + 2 \lambda_8},
    \ee
    leading to
    \be
    \label{v05}
    V_0 = V_0^{(5\pm)} \equiv
    - \frac{\left( \mu_2 \pm \mu_3 \right)^2}{2 \lambda_2
      + \lambda_4 + \lambda_6 + 2 \lambda_8}.
    \ee
  \item $\cos{\vartheta} = - \mu_3 \left/ \left( 4 \lambda_8 v_2 v_3 \right)
    \right.$.
    This leads to
    \be
    v_3^2 = v_2^2 = - \frac{\mu_2}{2 \lambda_2 + \lambda_4 + \lambda_6
      - 2 \lambda_8} \label{prop3}
    \ee
    and to
    \be
    \label{v06}
    V_0 = V_0^{(6)} \equiv - \frac{\mu_2^2}{2 \lambda_2 + \lambda_4 + \lambda_6
      - 2 \lambda_8} - \frac{\mu_3^2}{4 \lambda_8}.
    \ee
    Of course,
    this stability point only exists if $\left| \cos{\vartheta} \right| \le 1$,
    \textit{viz.}
    \be
    1 \le
    \left| \frac{4 \lambda_8 \mu_2}{\left( 2 \lambda_2 + \lambda_4 + \lambda_6
    - 2 \lambda_8 \right) \mu_3} \right|. \label{prop4}
    \ee
  \end{enumerate}
\end{enumerate}

\end{appendix}

\end{document}